\documentclass[11pt]{article}
\usepackage{amsfonts}
\usepackage{amssymb}
\usepackage{graphicx}
\usepackage{amsmath}
\usepackage{amsthm}
\usepackage{color}
\usepackage{multirow}
\usepackage{url}
\usepackage{float}
\usepackage{booktabs} 
\usepackage{makecell}
\usepackage{listings}
\usepackage{subcaption}
\usepackage{algorithm}
\usepackage{enumerate}
\usepackage{tikz}

\usepackage{algpseudocode}
\lstnewenvironment{Rcode}{\lstset{language=R}}{}

\newtheorem{theorem}{Theorem}[section]

\newtheorem{corollary}[theorem]{Corollary}

\newtheorem{proposition}[theorem]{Proposition}
\newtheorem{remark}[theorem]{Remark}

\definecolor{battleshipgrey}{rgb}{0.52, 0.52, 0.51}
\definecolor{navyblue}{rgb}{0.0, 0.0, 0.5}
\definecolor{arsenic}{rgb}{0.23, 0.27, 0.29}
\definecolor{oldmauve}{rgb}{0.4, 0.19, 0.28}
\usepackage[colorlinks=TRUE]{hyperref}
\hypersetup{
	colorlinks=true,
	linkcolor=navyblue,
	filecolor=blue,      
	urlcolor=oldmauve,
	citecolor=navyblue,
	pdftitle={Overleaf Example},
	pdfpagemode=FullScreen,
}
\usepackage{natbib}
\setcitestyle{authoryear}

\begin{document}

	\title{ \bf Fast and efficient algorithms for sparse semiparametric bi-functional regression}
	\author{Silvia Novo{$^a$}\footnote{Corresponding author email address: \href{s.novo@udc.es}{s.novo@udc.es}} \hspace{2pt} Philippe Vieu{$^b$} \hspace{2pt} Germ\'{a}n Aneiros{$^c$} \\		
		{\normalsize $^a$ Department of Mathematics, MODES, CITIC, Universidade da Coruña, A Coruña, Spain}\\
		{\normalsize $^b$ Institut de Math\'{e}matiques, Universit\'e  Paul Sabatier, Toulouse, France}\\
		{\normalsize $^c$ Department of Mathematics, MODES, CITIC, ITMATI, Universidade da Coruña, A Coruña, Spain}
	}
	
	\date{}
	\maketitle
	\begin{abstract} A new sparse semiparametric model is proposed, which incorporates the influence of two
		functional random variables in a scalar response in a flexible and interpretable manner.
		One of the functional covariates is included through a single-index structure, while the
		other is included linearly through the high-dimensional vector formed by its discretised
		observations. For this model, two new algorithms are presented for selecting relevant
		variables in the linear part and estimating the model. Both procedures utilise the functional
		origin of linear covariates. Finite sample experiments demonstrated the scope of application
		of both algorithms: the first method is a fast algorithm that provides a solution (without loss
		in predictive ability) for the significant computational time required by standard variable
		selection methods for estimating this model, and the second algorithm completes the set
		of relevant linear covariates provided by the first, thus improving its predictive efficiency.
		Some asymptotic results theoretically support both procedures. A real data application
		demonstrated the applicability of the presented methodology from a predictive perspective
		in terms of the interpretability of outputs and low computational cost.
	\end{abstract}
	
	
	\section{Introduction}
	
	\subsection{Towards semiparametric sparse regression modelling}
Owing to the technological advances in data storage and collection, variables are observed
to more frequently vary over a continuum (data include curves, images, etc.). This
informative richness provided by functional variables makes them useful in regression problems,
which could involve more than one functional object. To be adapted to such situations,
the main goal of regression modeling in the context of functional data analysis (FDA) involves
the development of models combining the flexibility and interpretability of derived
estimations, without being significantly sensitive to the effects of dimensionality. Recent trends in FDA (see e.g. \citealt*{goiv17}, and \citealt{anecfgv19} for general surveys) have highlighted the necessity of creating models and procedures that can reduce the dimensionality
of problems (see e.g. \citealt*{vie17} for a specific survey), and both semiparametric and sparse ideas
appear to be of interest for achieving this purpose.

	With respect to semiparametrics in FDA, partial linear ideas have been extended to
	FDA following different approaches (see e.g. \citealt*{aneiros_2006,aneiros_2011}, \citealt*{lian2011}, \citealt*{maity_2012}),  along with procedures for controlling dimensionality, such as variable
	selection techniques (see \citealt*{aneirosfv_2015}). To control adverse dimensionality
	effects, single-index structures were proposed in semiparametric FDA (the studies by \citealt*{ferpv03}, \citealt{ait} or \citealt*{novo_2019}, are related to the functional single index model (FSIM)). By combining partial linear and
	single-index structures, a functional partial linear single-index model was obtained (see \citealt*{wang_2016}), thus creating balance between flexibility and interpretability. Besides semiparametrics, sparse modelling ideas have been recently used in FDA to control
	the effects of dimensionality. For instance, the standard penalised least squares (PLS) method
	for variable selection has been extended to semiparametric FDA models (see e.g. \citealt*{aneirosfv_2015}, and \citealt*{novo}, and references therein).
	
	In this study we investigated a situation wherein multiple functional predictors are
	included in the statistical sample. Accordingly, we present a new model based on the combination
	of partial linear, single-index, and sparse ideas, called multifunctional partial linear
	single-index model (MFPLSIM); to ensure clarity, we focus on the bifunctional case. The
	main idea involves modelling the effects of each functional covariate in a different manner,
	i.e., one of the functional covariates ($\mathcal{X}$) enters the model through a semiparametric single-index continuous structure% ({\color{blue}{as in \citealt*{novo_2019} and \citealt*{novo}}})
	, while the other one ($\mathcal{\zeta}$) enters linearly through  the $p_n$-dimensional vector built from its discretised observations. 
	Morover, MFPLSIM incorporates the continuous
	and point-wise effects of functional variables, involving interpretable parameters in
	both cases. Furthermore, it should be considered that we have a significantly large number
	of linear covariates $p_n$  and only a few of them affect the response. Therefore, this flexibility
	must to be combined with an accurate variable selection method. However, application
	of the standard PLS method to MFPLSIM becomes significantly infeasible owing to the
	huge computational time required for variable selection even for moderate values of $p_n$. In addition, the standard procedures, originating from the adaptation of the multivariate
	methodology to FDA, do not consider the strong correlation structure present between linear
	covariates owing to its functional origin (although there exist some proposals in the statistical
	literature for selecting covariates in linear models with features that can be ordered
	in some meaningful way, such as group LASSO, see \citealt*{bakin} and fused LASSO, see \citealt*{tisa}; among others). Accordingly, we developed two new algorithms for variable
	selection in the linear part as well as model estimation, which utilises the functional origin
	of these scalar variables included in the linear part. In both algorithms, MFPLSIM will be
	transformed in a certain linear regression model wherein the correlation between covariates
	is attenuated, followed by the application of a standard PLS procedure. For this, the LASSO
	penalty (see \citealt*{tibshirani}) or the SCAD penalty (see \citealt*{fan_li}) could be considered. We used the SCAD penalty, which possesses the oracle property (a property not possessed
	by the LASSO). Other approaches with the oracle property include the adaptive LASSO (see \citealt*{zou}) and the bridge estimators when $0<\gamma<1$ is considered in the corresponding penalty function $\mathcal{P}_{\lambda,\gamma}\left(\beta\right)=\lambda\left|\beta\right|^{\gamma}$ (see \citealt*{huahowa}).

	\subsection{Applied issues on chemometrics}\label{chemometrics}
	Chemometrics is a field of applied sciences in which FDA is employed (see \citealt*{fer2006} and references therein). For instance, to analyse and/or detect some components of a chemical mixture, the
	spectrometric data obtained by measuring the light absorbance of the mixture for several
	different wavelengths are commonly observed. Hence, long, expensive (and occasionally
	dangerous) chemical experiments can be avoided by analysing the spectrometric data. Let
	us look at a specific example.
	
\begin{figure}[htp]
		\centering
		\makebox{\includegraphics[width=0.33\textwidth]{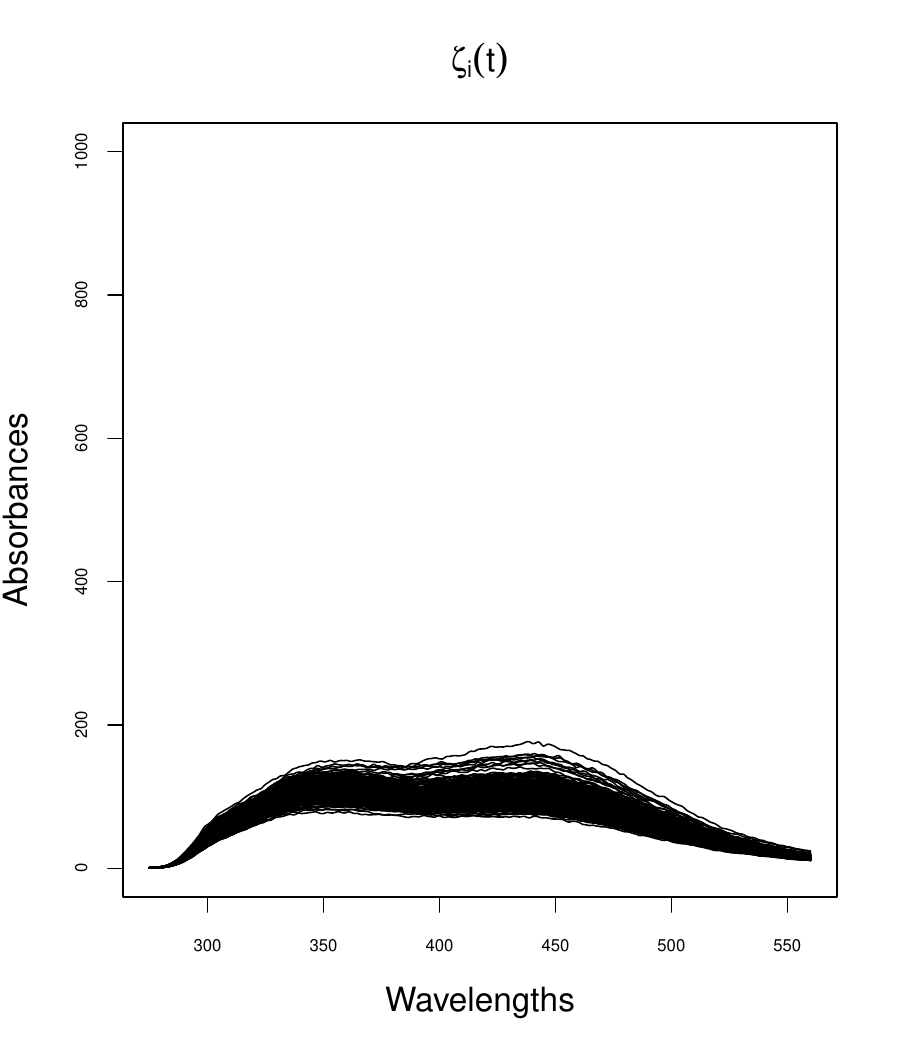}
			\includegraphics[width=0.33\textwidth]{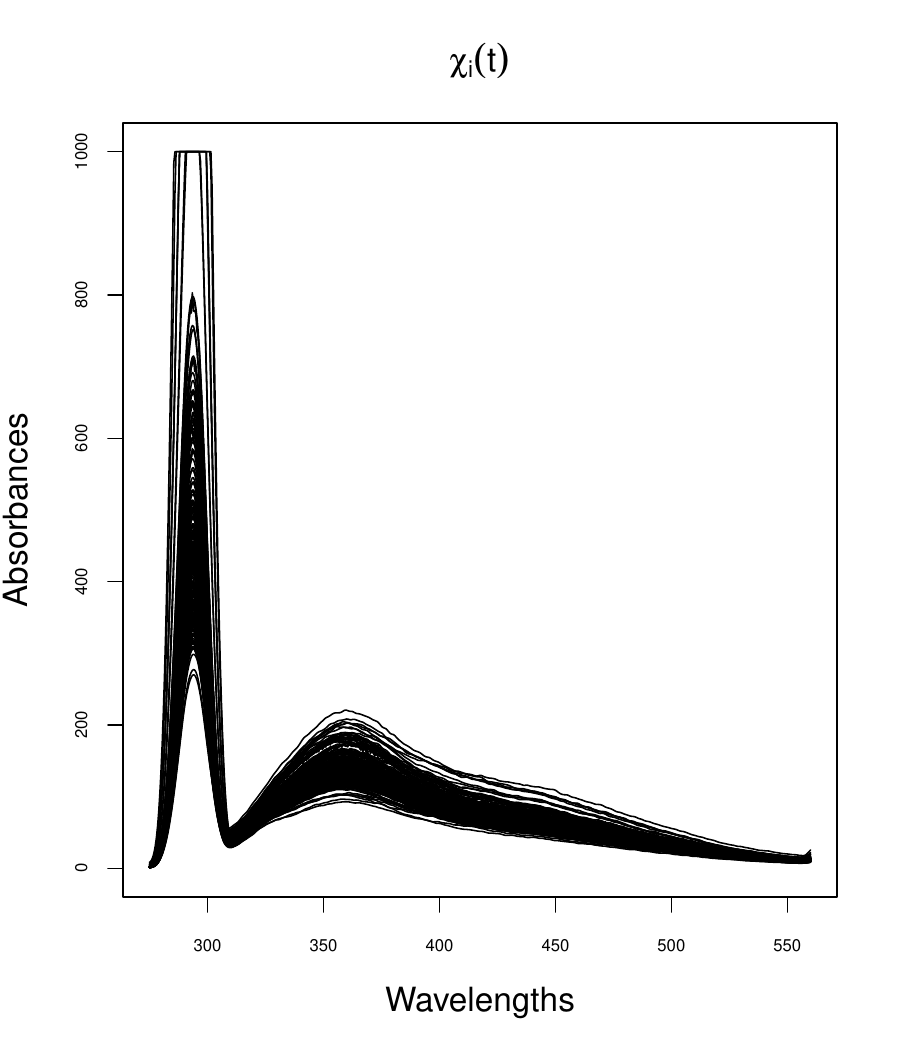}}
		\caption{\label{fig3} Left panel:  Absorbance curves at excitation wavelengths  $240nm$ ($\mathcal{\zeta}$). Right panel: Absorbance curves at excitation wavelengths $290nm$ ($\mathcal{X}$).}
		
	\end{figure}

	At a sugar plant in Scandinavia, $268$ samples were obtained by sampling sugar every $8h$ over $3$ months. For each sample, the absorbance spectra from $275$ to $560nm$ were measured at an interval of $0.5nm$ (therefore, $p_n=571$) and excitation wavelengths $240$  and $290nm$ (denoted by $\mathcal{\zeta}$ and $\mathcal{X}$, respectively). Samples of both curves can be seen in Figure \ref{fig3}.
	%{\color{red}{Aqui tambien los graficos estan bonitos en color pero lo mas ciertoe s que sean publicados en blanco y negre}}

	In the experiment, the ash content, $Y$, was also determined for each sample; it remains
	to be investigated whether the value of ash content for some new sample can be predicted
	by observing its spectrometric curves. This is a typical regression problem with scalar response (the ash content $Y$). At this stage, there are two alternative approaches for analysing the
	spectrometric data:\\
	
	\noindent i) Consider the data as $p_n$-dimensional vectors composed of the values observed at discretised
	wavelengths;\\
	
	\noindent ii) Consider the data as curves obtained by smoothing discretised observations.\\
	
		Both
	approaches have advantages and/or drawbacks, but both tackle the same dimensionality
	problem.
	The first option leads to multivariate regression analysis where the number of variables ($2p_n=1142$ variables) is significantly higher than the sample size ($n=268$) and the
	correlation between variables is high. When dealing with this discretised perspective, the
	techniques used in big data analysis, such as sparse modelling, must be suitably adapted. However, the second option involves only two predictors (namely, the curves $\mathcal{\zeta}$ and $\mathcal{X}$) but
	both are elements of infinite dimensional functional space. When dealing with this continuous
	point of view, regression techniques that are insensitive to the dimensionality of covariates,
	such as semiparametric modelling, should be suitably adapted.
	
Given the complexity of the sample, such spectrometric data should be analysed in a
flexible manner. The methodology developed in this study achieves this goal by developing
a model that combines the first and second options and providing statistical methods that
combine both sparse and semiparametric concepts.

	\subsection{Presentation of the paper}

	The remainder of this paper is organised as follows. In Section 2, MFPLSIM is presented
	focusing on the sparse feature of the model. Section 3 discusses the first variable selection
	method presented by us, i.e., a fast algorithm that ensures variable selection and estimation
	of MFPLSIM in appropriate time, even for significantly large values of $p_n$. Moreover, a
	simulation study was performed that demonstrated the practical advantages of MFPLSIM
	(computational time and no loss in predictive efficiency) in comparison to the standard PLS
	method. In Section 4, we present a second variable selection procedure, a more refined
	algorithm that specifies and completes the set of relevant variables. Finite sample simulated
	experiments enabled us to compare its performance with that of the first algorithm. Thus,
	we infer the scope of application of each variable selection procedure. Finally, the real data
	application presented in Section 2 is analysed using the proposed algorithms, thus illustrating
	the advantages of the methodology, i.e., high predictive power, interpretable outputs,
	and reasonably low computational time. Accordingly, a wide scope of asymptotics can be
	obtained for the mathematical assessment of the procedure. To facilitate comprehension,
	these theoretical issues have been relegated to the Appendix.

	\section{Proposed model}
	
In this study, we consider a regression problem with scalar response $Y$ and two functional predictors $\zeta$ and $\mathcal{X}$. To increase the flexibility of the model, both functional covariates
are not assumed to act on the response similarly. Precisely, MFPLSIM is defined by
assuming that $\zeta$ acts only through its $p_n$ discretised points ($n$ denotes the sample size) while $\mathcal{X}$ acts in a continuous semiparametric manner, leading to the following regression relationship:  
	\begin{equation}
		\label{model}
		Y=\sum_{j=1}^{p_n}\beta_{0j}\zeta(t_j)+m\left(\left<\theta_0,\mathcal{X}\right>\right)+\varepsilon, 
	\end{equation}
	where:
	\begin{itemize}
		\item $Y$ is a real random response, and $\mathcal{X}$ denotes a random element belonging to some separable Hilbert space $\mathcal{H}$ with inner product denoted by $\left<\cdot,\cdot\right>$. The second functional predictor $\zeta$ is supposed to be a random curve defined on some interval $[a,b]$ which  is observed at points $a\leq t_1<\dots<t_{p_n}\leq b$. In addition, a single index  of $\mathcal{X}$ is sufficient to summarise all the information carried in $\mathcal{X}$ to predict $Y$ and $\zeta(t_j)$ ($j=1,\ldots,p_n$); that is,$$\textrm{E}\left(Y|\zeta(t_1),\dots,\zeta(t_{p_n}),\mathcal{X}\right)=\textrm{E}\left(Y|\zeta(t_1),\ldots,\zeta(t_{p_n}),\left<\theta_0,\mathcal{X}\right>\right)$$ and $$\textrm{E}\left(\zeta(t_j)|\mathcal{X}\right)=\textrm{E}\left(\zeta(t_j)|\left<\theta_0,\mathcal{X}\right>\right), \ j=1,\ldots,p_n.$$
		(Note that the notation $\left<\cdot,\cdot\right>$ denotes any inner product)
		\item  $\left(\beta_{01},\dots,\beta_{0p_n}\right)^{\top}$ is a vector of unknown real coefficients and $m$ denotes a smooth unknown link function. In addition, $\theta_0$ is an unknown functional direction in $\mathcal{H}$.  
		\item $\varepsilon$ denotes the random error.
	
	\end{itemize}

	Next, we wish to include a feature in model (\ref{model}) that considers the case in which only a
	few among the discretised points $\zeta(t_j)$ affect the response $Y$. In other words, only a few scalar linear variables among the set $\{\zeta(t_1),\dots,\zeta(t_{p_n})\}$ have to be part of the model. Mathematically,
	this can be modelled using sparse regression ideas that define the set of significant indices as
	\begin{equation*}
		S_n=\{j=1,\dots,p_n, \quad \textrm{such that } \beta_{0j}\not= 0\}, 
	\end{equation*}
	and by assuming standard conditions such as
	\begin{equation} \sharp S_n=s_n=o(p_n),\label{number_imp_points}
	\end{equation}
	\begin{equation}\textrm{exists } c,\ \ \textrm{for all } n,\ \ \textrm{such that } \sum_{j\in S_n}\left|\beta_{0j}\right|<c<\infty.\label{iv2}\end{equation}
	
	To ensure the identifiability of the model (\ref{model}), we assume that either $\left<\Gamma \theta_0,\theta_0\right>=1$, where $\Gamma$ denotes the covariance function of the functional variable $\mathcal{X}$ and $(\Gamma \theta_0)(t)=\left<\Gamma(\cdot,t),\theta_0\right>$, or $\left<\theta_0,\theta_0\right>=1$, and that for some arbitrary $t_0$ in the domain of $\theta_0$, one has $\theta_0(t_0)>0$. These conditions are common in literature of semiparametric models (\citealt{ait} or \citealt*{wang_2016}).
	
	In the literature, model (\ref{model}) has been studied for cases in which the covariates of the
	linear and semiparametric component are finite-multidimensional, such as the partially linear
	single-index model (PLSIM) introduced in \cite{carroll_1997}. For this model, \cite{liang_2010} studied the variable selection problem over the set of fixed $p$ linear covariates using the PLS approach. Recently, multivariate concepts have been extended to FDA. Accordingly,
	\cite*{wang_2016} presented a semi-functional partial linear single-index model
	(SFPLSIM), where a functional variable enters the single-index component, while the linear
	part of the model is finite-multidimensional. In addition, \citealt*{novo} studied the sparse semi-functional partial linear single-index model (SSFPLSIM), a generalisation of the SFPLSIM in a sparse context.  %To the best of our knowledge, nowadays there are no work dealing with variable selection in the SFPLSIM model.
	
	We attempt to study model (\ref{model}) in which the  linear components are obtained from the
	discretization of $\zeta$. At this stage, it is worth noting that this cannot be performed as a direct
	application of earlier methodologies. This is because the variables $\zeta(t_j)$ are obtained from
	a continuous variable, thereby adding the following two main methodological difficulties
	in the estimation and variable selection tasks. The continuous nature of $\zeta$ causes strong correlation between between variables with linear effect, i.e., when $t_j$ is close from $t_k$, the two respective variables $\zeta(t_j)$ and $\zeta(t_k)$ roughly contain the same information on the response $Y$. In addition,
	in various applications, $p_n$ is often significantly large number, leading to high-dimensional
	problems. This should be considered with the estimation of direction $\theta_0$ which usually has
	a high computational cost. Thus, it is crucial to develop specific tools for selecting relevant
	variables and estimating model (\ref{model}) in a feasible computational time.

	\section{FASSMR algorithm}
	\label{pvs 1 step}
In practice, the progress in measurement technologies leads to numerous discretisations
of the functional variable $\zeta$ in many situations. It is well known that for any standard
variable selection method (such as PLS), for larger values of $p_n$ more computational time is required. Therefore, to obtain results in a reasonable amount of time, it is important to
develop algorithms that reduce the computational time.
	
It should be noted that in the multivariate case, more variables in the linear part generally
indicate varied external information about the response; by contrast, when linear covariates
have a functional origin, with bigger $p_n$ we obtain more precise information about the single continuous process that results in a discretized curve $\zeta$.
	Accordingly, we propose the
	following fast algorithm for sparse semiparametric multifunctional regression (FASSMR).
	This algorithm considers a reduced model with few linear covariates (but covering the entire
	discretization interval of $\zeta$), and directly discards the other linear covariates (they contain
	similar information about the response).

	\subsection{Procedure}
To introduce the variable selection algorithm,  we assume a statistical sample of size $n$:
	\begin{equation}
		\left\{(\zeta_i,\mathcal{X}_i,Y_i),\quad i=1,\dots,n\right\}\ \ \textrm{i.i.d.  to  } \ \ (\zeta,\mathcal{X},Y). \label{sample}
	\end{equation} 
	We assume, without lost of generality, that the number of linear covariates $p_n$ can be expressed as follows: $p_n=q_nw_n$ with $q_n$ and $w_n$ integers.
	The previous consideration allows us to present a subset of the initial $p_n$ linear covariates, which contains only $w_n$ equally spaced discretised observations of $\zeta$, covering the whole interval $[a,b]$. This subset can be given as follows:
	\begin{equation}
		\mathcal{R}_n^{\pmb{1}}=\left\{\zeta\left(t_k^{\pmb{1}}\right),\ \ k=1,\dots,w_n\right\},\label{initial_set}
	\end{equation} 
	where $t_k^{\pmb{1}}=t_{\left[(2k-1)q_n/2\right]}$ and $\left[z\right]$ denotes the smallest integer not less than $z\in\mathbb{R}$. 
	
	It is noteworthy that the correlation between consecutive variables within $\mathcal{R}_n^{\pmb{1}}$ is considerably less important than that in the entire set of $p_n$ initial linear covariates. Therefore, one may reasonably expect that the behaviour of the standard PLS method (see \citealt*{novo}) will be better if it is applied between the variables in $\mathcal{R}_n^{\pmb{1}}$ instead of using the entire set of $p_n$ linear covariates; moreover, we expect a significantly reduced computational
	time owing to the use of moderate values for $w_n$.
Consequently, the standard PLS variable
selection procedure was applied between the variables in 
	$\mathcal{R}_n^{\pmb{1}}$.
	
	Hence, we consider the following reduced model, which only involves the linear covariates
	belonging to $\mathcal{R}_n^{\pmb{1}}$:
	\begin{equation}
		Y_i=\sum_{k=1}^{w_n}\beta_{0k}^{\pmb{1}}\zeta_i(t_k^{\pmb{1}})+m^{\pmb{1}}\left(\left<\theta_0^{\pmb{1}},\mathcal{X}_i\right>\right)+\varepsilon_i^{\pmb{1}}.
		\label{mod_red}
	\end{equation}
	For this model, the set of relevant indices and its estimation can be denoted by
	\begin{equation*}
		\mathcal{S}_{n}^{{\pmb{1}}}=\{k=1,\dots,w_n,\quad \beta_{0k}^{\pmb{1}}\not=0\},
	\end{equation*}
	\begin{equation*}
		\widehat{\mathcal{S}}_{n}^{\pmb{1}}=\{k=1,\dots,w_n, \quad \widehat{\beta}_{0k}^{\pmb{1}}\not=0\},
	\end{equation*}
	with $s_{n}^{\pmb{1}}=\sharp(\mathcal{S}_{n}^{\pmb{1}})$ and $\widehat{s}_{n}^{\pmb{1}}=\sharp\left(\widehat{\mathcal{S}}_{n}^{\pmb{1}}\right)$.
	In addition, it is assumed that %at least one of the $w_n$ linear coefficients of (\ref{mod_red}) is non-null:
	\begin{equation}\textrm{exists } c,\ \ \textrm{for all } n,\ \ \textrm{such that} \ \ \inf_{n}\min_{k\in\mathcal{S}_{n}^{\pmb{1}}}\left|\beta_{0k}^{\pmb{1}}\right|>c>0.\label{iv3}\end{equation}

	Then, the variable selection task can be developed with the following steps.
	\begin{enumerate}
		\item First, model (\ref{mod_red}) is transformed into a linear model by extracting the effect of functional
		variable $\mathcal{X}_i$  (when it is projected along the direction $\theta_0^{\pmb{1}}$) from $Y_i$ and $\zeta_i(t_k^{\pmb{1}})$ ($k=1,\dots,w_n$). Specifically,
		\begin{equation}
			Y_i-\textrm{E}\left(Y_i|\left<\theta_0^{\pmb{1}},\mathcal{X}_i\right>\right)=\sum_{k=1}^{w_n}\beta_{0k}^{\pmb{1}}\left(\zeta_i(t_k^{\pmb{1}})-\textrm{E}\left(\zeta_i(t_k^{\pmb{1}})|\left<\theta_0^{\pmb{1}},\mathcal{X}_i\right>\right)\right)+\varepsilon_i^{\pmb{1}}.
			\label{lin_mod}
		\end{equation}
		Because the conditional expectations in expression (\ref{lin_mod})  are unknown, they may be estimated via regression. Nadaraya--Watson-type estimators are used for estimating these regressions. Therefore, we consider the following $n\times n$-matrix of local weights:
		\begin{equation}
			\pmb{W}_{h,\theta}=\left(w_{n,h,\theta}(\mathcal{X}_i,\mathcal{X}_{\ell})\right)_{i,\ell=1,\dots,n},\ \textrm{ with } \
			w_{n,h,\theta}(\chi,\mathcal{X}_{i})=\frac{K\left(d_{\theta}\left(\chi,\mathcal{X}_{i}\right)/h\right)}{\sum_{\ell=1}^{n} K\left(d_{\theta}\left(\chi,\mathcal{X}_{\ell}\right)/h\right)},  \label{local_weights}
		\end{equation}
		where $h>0$ denotes the bandwidth, $K$ is the kernel and $d_{\theta}(\cdot,\cdot)$ is the semimetric defined as
		\begin{equation*}
			d_{\theta}(\chi,\chi')=\left|\left<\theta,\chi-\chi'\right>\right|, 
		\end{equation*}
		for each $\chi,\chi',\theta\in\mathcal{H}$, which measures the proximity between projected curves on direction $\theta$. 
		Consequently, we obtain the following transformed variables for each $\theta\in\mathcal{H}$:
		\begin{equation*}
			\widetilde{\pmb{Y}}_{\theta}=\left(\pmb{I}-\pmb{W}_{h,\theta}\right)\pmb{Y},\quad \widetilde{\pmb{\zeta}}_{\theta}^{\pmb{1}}=\left(\pmb{I}-\pmb{W}_{h,\theta}\right)\pmb{\zeta}^{\pmb{1}},
		\end{equation*}
		where $\pmb{\zeta}^{\pmb{1}}$ is the $n\times w_n$ matrix $(\zeta_i(t_k^{\pmb{1}}), 1\leq i\leq n, 1\leq k\leq w_n)$, and $\pmb{Y}$ denotes the respective vector of responses $(Y_1,\dots,Y_{n})^{\top}$. 
		\item The standard PLS variable selection procedure is applied in the set $\mathcal{R}_n^{\pmb{1}}$. Specifically, the penalised profile least squares function is minimised over the pair $(\pmb{\beta}^{\pmb{1}},\theta^{\pmb{1}})$ with $\pmb{\beta}^{\pmb{1}}\in\mathbb{R}^{w_n}$ and  $\theta^{\pmb{1}}\in \Theta_n^{\pmb{1}}\subset\mathcal{H}$:
		\begin{equation}
			\mathcal{Q}^{\pmb{1}}\left(\pmb{\beta}^{\pmb{1}},\theta^{\pmb{1}}\right)=\frac{1}{2}\left(\widetilde{\pmb{Y}}_{\theta^{\pmb{1}}}-\widetilde{\pmb{\zeta}}_{\theta^{\pmb{1}}}^{\pmb{1}}\pmb{\beta}^{\pmb{1}}\right)^{\top}\left(\widetilde{\pmb{Y}}_{\theta^{\pmb{1}}}-\widetilde{\pmb{\zeta}}_{\theta^{\pmb{1}}}^{\pmb{1}}\pmb{\beta}^{\pmb{1}}\right)+n\sum_{k=1}^{w_n}\mathcal{P}_{\lambda_{k_n}}\left(|\beta_{k}^{\pmb{1}}|\right),
			\label{func_min1}
		\end{equation}
		where $\mathcal{P}_{\lambda_{k_n}}\left(\cdot\right)$ is the SCAD-penalty function, which is defined for $a>2$ as follows:
		\begin{equation}
			\mathcal{P}_{\lambda}=\left\{\begin{aligned}
				&\lambda\left|u\right|&  \quad |u|<\lambda,\\
				&\frac{(a^2-1)\lambda^2-(|u|-a\lambda)^2}{2(a-1)}&\quad \lambda\leq |u|<a\lambda, \\
				&\frac{(a+1)\lambda^2}{2}&\quad |u|\geq a\lambda.
			\end{aligned}
			\right.
			\label{SCAD}
		\end{equation} 
		The value $a=3.7$ is usually considered in literature  (see \citealt*{fan_li}).
		\item  $(\widehat{\pmb{\beta}}_0^{\pmb{1}},\widehat{\theta}_0^{\pmb{1}})$ denotes a local minimiser of the criterion $\mathcal{Q}^{\pmb{1}}(\cdot,\cdot)$, where $\widehat{\pmb{\beta}}_0^{\pmb{1}}=(\widehat{\beta}_{01}^{\pmb{1}},\dots,\widehat{\beta}_{0w_n}^{\pmb{1}})^{\top}$. Then, $\zeta(t_k^{\pmb{1}})$ is selected in $\mathcal{R}_n^{\pmb{1}}$ if, and only if, $\widehat{\beta}_{0k}^{\pmb{1}}\not =0$.
	\end{enumerate}
	
	\begin{remark} As expected, to obtain asymptotic results related to the presented variable selection algorithm (FASSMR), two types of assumptions should be considered. First, specific assumptions to treat covariates with the linear effect obtained from the discretisation of a
		curve (functional nature of linear covariates). Second, general assumptions to deal with the
		standard PLS procedure. Both types of assumptions are presented in the Appendix (see
		forthcoming conditions (\ref{dis})--(\ref{s1}) and \ref{sfplsim1})--(\ref{oracle_scad1}), respectively). It is worth noting that the
		assumptions related to the standard PLS procedure (conditions (\ref{sfplsim1}) and (\ref{oracle_scad1})) are written in a
		rather general form, such that they could be obtained from different sets of assumptions. For instance, \cite*{novo} present assumptions under which (\ref{sfplsim1}) and (\ref{oracle_scad1}) hold. In addition, in \cite*{novo}, (i) the existence of a local minimiser, $(\widehat{\pmb{\beta}}_0^{\pmb{1}},\widehat{\theta}_0^{\pmb{1}})$, of $\mathcal{Q}^{\pmb{1}}(\cdot,\cdot)$ was proven, (ii) the corresponding convergence rates are obtained, and (iii) the subset of eligible directions, $\Theta_n^{\pmb{1}}$, was theoretically characterised (practical considerations about $\Theta_n^{\pmb{1}}$ are included in Subsection \ref{T_C}). In addition, they included specific requirements for
		a general penalty function $\mathcal{P}_{\lambda}$, which were satisfied by the SCAD penalty used in this study. Finally, it is also worth noting that when dealing with partial linear single-index models, additional assumptions are usually considered to ensure identifiability. Such assumptions link the two types of covariates
		in the model ($\zeta$ and $\mathcal{X}$ in the case of the proposed MFPLSIM) and prevent the possibility of covariates with different types of effect (linear and semiparametric effects) being equal (see condition (vi) in \citealt*{liang_2010} and condition (25) in \citealt*{novo} for the cases of scalar and functional covariates, respectively). In this study, condition (25) in \citealt*{novo} is implicitly assumed (the best of our knowledge, \citealt*{novo} is the only paper in the statistical literature that deals with variable selection using the standard PLS procedure in SSFPLSIMs).

		\label{remark_t}
	\end{remark}
	\subsection{Outputs of FASSMR}
	After applying the variable selection procedure, the parameters of the model can be estimated. 
	Consequently, returning to model (\ref{model}) and considering the entire set of initial of $p_n$ linear covariates, a variable $\zeta(t_j)\in\{\zeta(t_1),\dots,\zeta(t_{p_n})\}$ is selected if, and only if, it belongs to $\mathcal{R}_n^{\pmb{1}}$ and its estimated coefficient, which can be denoted by $\widehat{\beta}_{0k_j}^{\pmb{1}}$,  is non-null. 
	Therefore, the following estimated set of relevant variables is obtained:
	\begin{equation*}
		\widehat{S}_n=\left\{j=1,\dots,p_n,\quad \textrm{such that } t_j=t_{k_j}^{\pmb{1}} \textrm{ with } \zeta(t_{k_j}^{\pmb{1}})\in\mathcal{R}_n^{\pmb{1}} \textrm{ and } \widehat{\beta}_{0k_j}^{\pmb{1}}\not=0\right\}.
	\end{equation*}
	An estimator for the linear coefficients and $\theta_0$ can be naturally obtained using the estimations involved in the variable selection. Therefore,
	\begin{eqnarray}
		\widehat{\beta}_{0j}&=&\left\{ 
		\begin{aligned} \widehat{\beta}_{0k_j}^{\pmb{1}}\quad  \textrm{ if } j\in \widehat{\mathcal{S}}_{n},\nonumber\\
			0 \quad \textrm{ otherwise,}\nonumber\\
		\end{aligned}\right.\\
		\quad \quad
		\widehat{\theta}_0&=&\widehat{\theta}_0^{\pmb{1}}.  \nonumber                  
	\end{eqnarray}

	Finally, denoting the vector of estimated parameters by $\widehat{\pmb{\beta}}_0$, an estimator for the function $m_{\theta_0}(\cdot)\equiv m(\left<\theta_0,\chi\right>)$ can be obtained by smoothing the residuals of the parametric fit:
	\begin{equation}
		\widehat{m}_{\widehat{\theta}_0}(\chi)\equiv\widehat{m}\left(\left<\widehat{\theta}_0,\chi\right>\right)=
		\frac{\sum_{i=1}^n\left(Y_i-\pmb{\zeta}_i^{\top}\widehat{\pmb{\beta}}_0\right) K\left(d_{\widehat{\theta}_0}\left(\chi,\mathcal{X}_i\right)/h\right) }{\sum_{i=1}^nK\left(d_{\widehat{\theta}_0}\left(\chi,\mathcal{X}_i\right)/h\right)},\label{est_m}
	\end{equation}
	where we have denoted $\pmb{\zeta}_i=\left(\zeta_i(t_1),\dots,\zeta_i(t_{p_n})\right)^{\top}$. Note that the estimation of $m_{\theta_0}(\cdot)$ is obtained for $m_{\theta_0^{\pmb{1}}}^{\pmb{1}}(\cdot)$. In other words, $\widehat{m}_{\widehat{\theta}_0}(\chi)=\widehat{m}_{\widehat{\theta}_0^{\pmb{1}}}^{\pmb{1}}(\chi)$.

	\begin{remark} Now, we make some comments regarding the design of the points ($t_j$, $j=1,\ldots,p_n$)  
		over which the curve $\zeta$ is discretised along with the theoretical complexity of the algorithm. To simplify the presentation, the design was assumed to be an equispaced grid. This is
		not restrictive in practice because if the data are unbalanced, each observed curve can be
		smoothed (in the first stage) and then computed at certain regularly spaced points to create
		a new (balanced) curves dataset. It should be noted that our results remain the same even if
		the assumed equispaced grid is replaced by a regular grid, $a\leq t_1<\dots<t_{p_n}\leq b$, supposed to be regular in the sense: the exist $c_1, c_2$ such that for all $j=1,\ldots,p_n-1, \ 0<c_1p_n^{-1}<t_{j+1}-t_j<c_2p_n^{-1}<\infty$. To obtain the theoretical complexity of the algorithm, the following should be considered: (i) the construction linear model on which the variable selection procedure will be applied (that is, the estimate of the conditional expectations in (\ref{lin_mod})), and (ii) the application of the variable selection procedure to such linear model. For a fixed value $\theta \in \Theta_n^{\pmb{1}}$ (for the definition of $\Theta_n^{\pmb{1}}$, see Remark \ref{remark_t}), and given tuning parameters $h$, $w_n$ and $\lambda$, the theoretical complexity for (i) is $O(n^2w_n)$, while for (ii) the theoretical complexity of the more computationally efficient algorithm is $O(nw_n)$ (see \citealt*{shi}). Therefore, the theoretical complexity of the proposed FASSMR algorithm is $O(n^2w_n\sharp\Theta_n^{\pmb{1}})$. Moreover, for the standard PLS procedure (see \citealt*{novo}), the complexity is $O(n^2p_n\sharp\Theta_n)$ ($\Theta_n$ is the set of eligible directions $\theta$ when estimating the full model (\ref{model}); usually, $\Theta_n=\Theta_n^{\pmb{1}}$); therefore, it is expected that, in practice, in cases where $w_n\ll p_n$, our algorithm will be significantly faster than the
		standard one (this will be even more evident in situations where $w_n\ll n\ll p_n$).  Finally, it
		should be noted that the factor $n^2$ that appears in the aforementioned orders is a consequence
		of the complexity of the model than that of the algorithm (specifically, it is a consequence
		of the presence of the nonparametric component $m(\cdot)$; that is, if $m(\cdot)$ were known, $n^2$ should be replaced by $n$).
		
		\label{remark_tt}
	\end{remark}

	\subsection{Simulation study}\label{S1}
	In this section, we demonstrate how FASSMR achieves good performance with significantly
	reduced computational cost (in comparison to the standard procedure) through
	numerous simulated samples. In Section \ref{des}, we introduce the model on which the simulation
	is based. To ensure high degree of generality, the model involves a combination of
	smooth functional covariates and rough ones (Brownian motions). Then, in Section \ref{T_C} we discuss some practical issues linked with selecting the parameters of the method (with
	specific attention to the problem of selecting the key parameter $w_n$).
	Finally, the results are reported in Section \ref{sim1-res} wherein the computational time and quality of estimation are
	quantified and computed for FASSMR and the standard PLS procedure.
	
	\subsubsection{Design}\label{des}
	
\begin{figure}[htp]
		\centering
		\makebox{\includegraphics[width=0.33\textwidth]{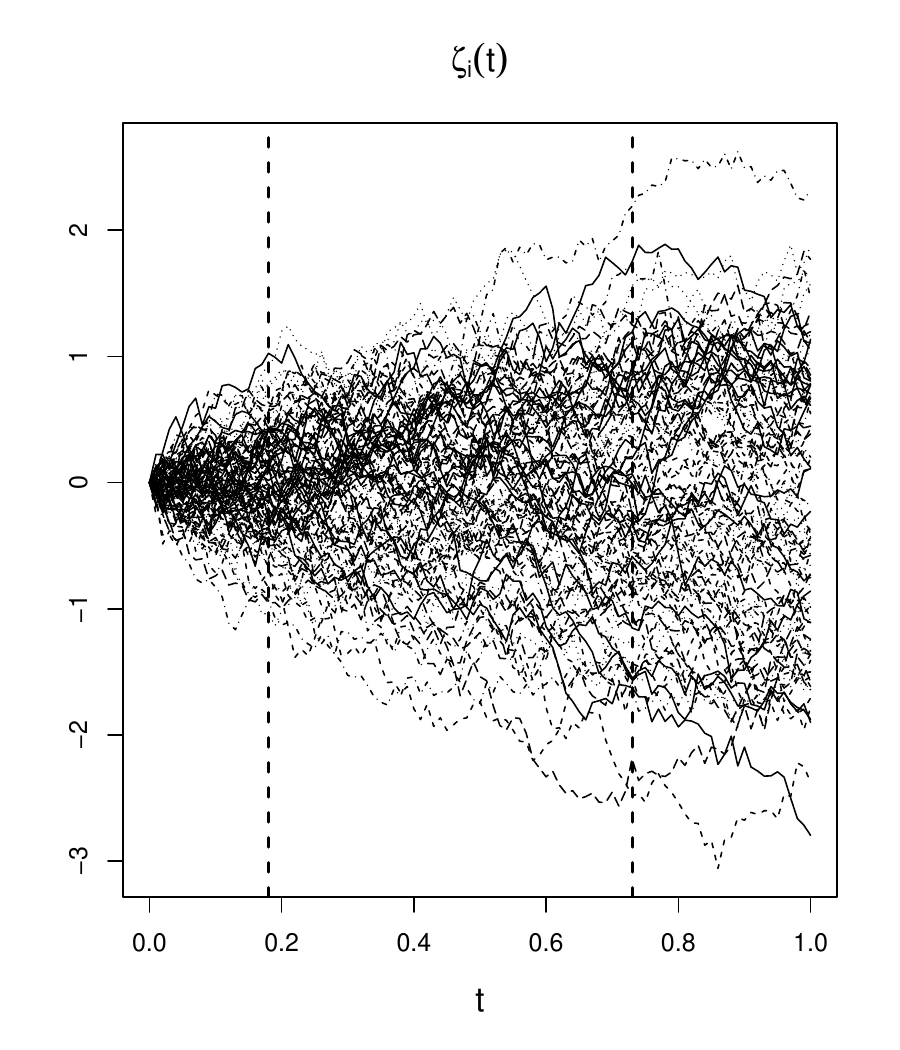}
			\includegraphics[width=0.33\textwidth]{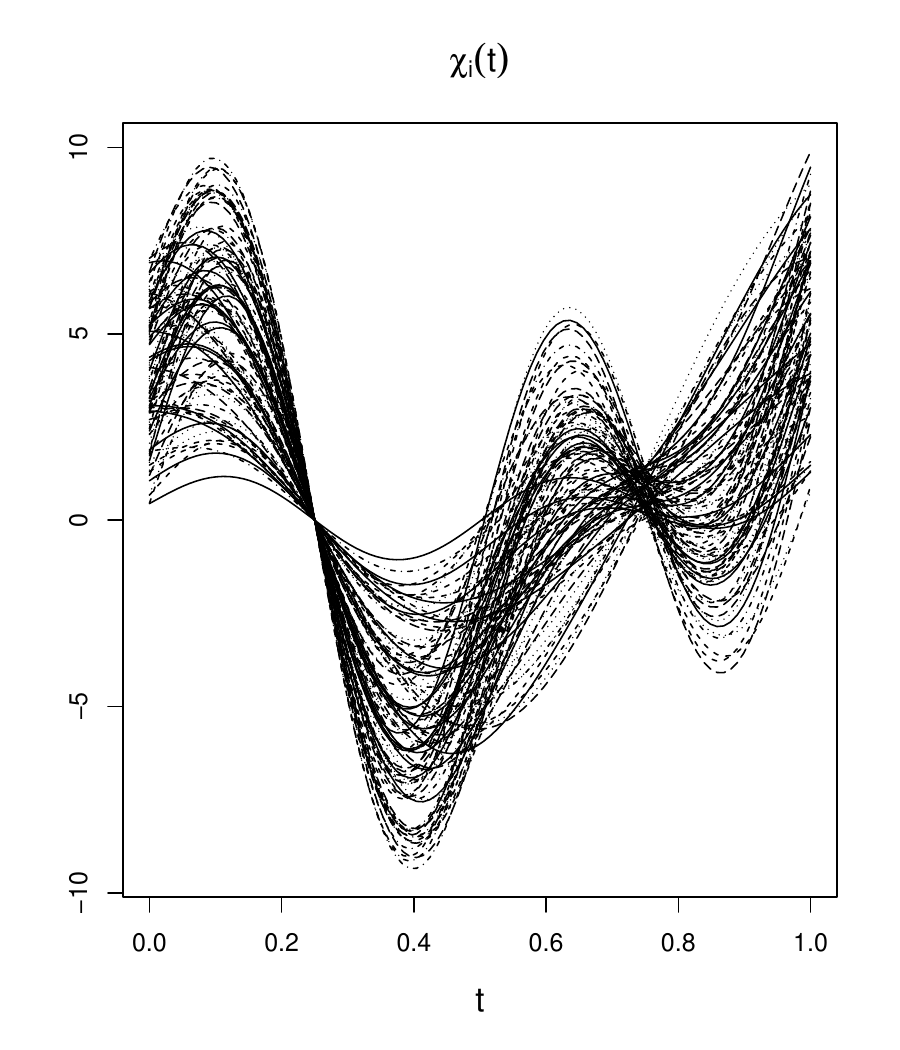}
			\includegraphics[width=0.33\textwidth]{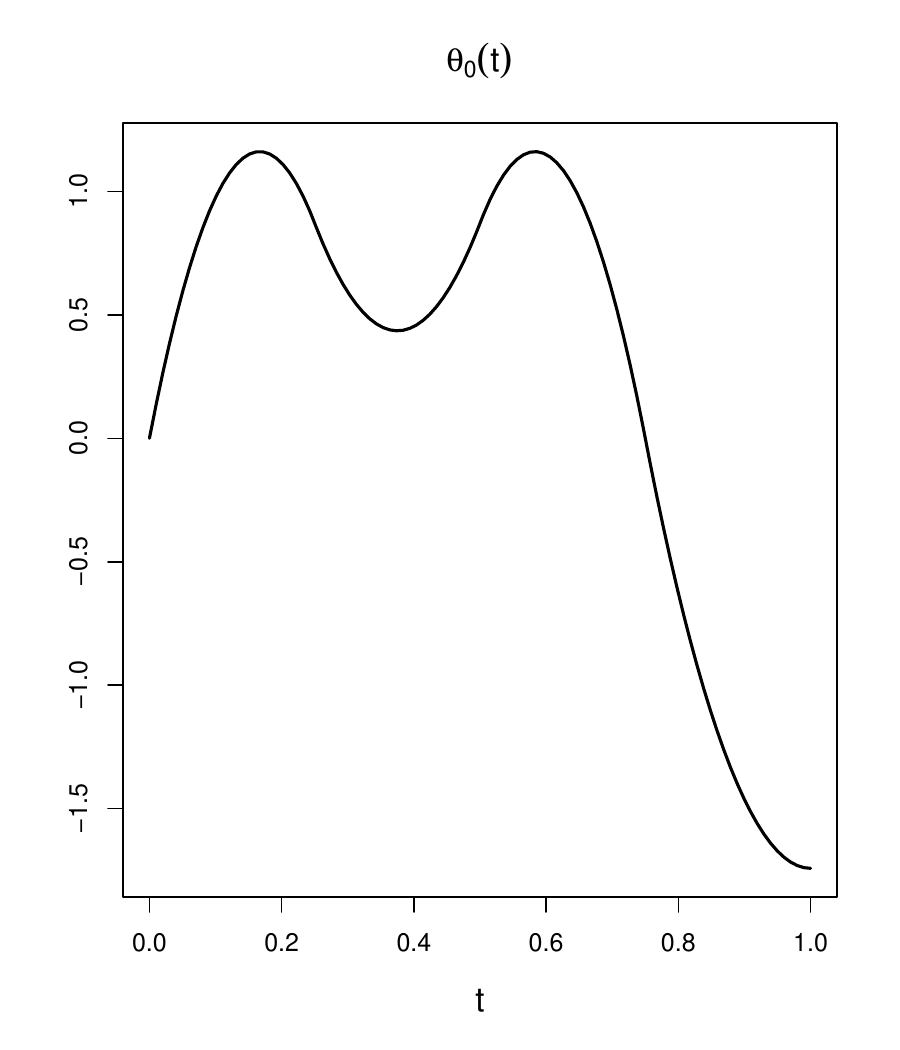}}
		\caption{\label{fig1} Right panel: $100$ sample paths of standard Brownian motion; dotted vertical lines mark the impact points at instants $t_{j1}=0.18$  and $t_{j2}=0.73$. Middle panel: sample of $100$ curves obtained from (\ref{X-sim}). Left panel: $\theta_0$.}	
	\end{figure}

	For different values of the sample size, $n\in\{100,200,300\}$, and different numbers of linear covariates, $p_n\in\{101,201,501,1001,10001\}$, we generated  observations i.i.d.  $\mathcal{D}=\{(\zeta_i,\mathcal{X}_i,Y_i),\quad i=1,\dots,n+100\}$ from the model:
	\begin{equation}
		Y_i=\sum_{j=1}^{p_n}\beta_{0j}\zeta_i(t_j)+m\left(\left<\theta_0,\mathcal{X}_i\right>\right)+\varepsilon_i, \label{mod_sim}
	\end{equation}
	where: 
	\begin{itemize}
		\item ${t_j}$ denotes equispaced points in $[0,1]$, with $t_1=0$ and $t_{p_n}=1$.
		\item $\zeta_i$ is a standard Brownian motion. We only consider two non-null coefficients
		$\beta_{0j_1}=2$ and $\beta_{0j_2}=-3$, with impact points $t_{j1}=0.18$  and $t_{j2}=0.73$ (left panel in Figure \ref{fig1} shows $100$ sample paths of the standard Brownian motion with influential points marked in dotted vertical lines).
		\item The curves involved in the nonlinear part were generated from:
		\begin{equation}
			\mathcal{X}_i(t)=a_i\cos(2\pi t) + b_i\sin(4\pi t) + 2c_i(t-0.25)(t-0.5) \ \ \ \textrm{for all } t \in [0,1],
			\label{X-sim}
		\end{equation}
		where the random variables $a_i$, $b_i$ and $c_i$  ($i=1,\dots,n+100$) are independent (both between and within vectors $(a_i,b_i,c_i)^{\top}$) and uniformly distributed on the interval $[0,6]$. These curves were discretised on the same grid of $100$ equispaced points in $[0,1]$ (representation of a sample of $100$ curves can be seen in the middle panel of Figure \ref{fig1}).
		\item The true direction of projection was generated using a B-splines basis, i.e., 
		\begin{equation}
			\theta_0(\cdot)=\sum_{j=1}^{d_n}\alpha_{0j}e_j(\cdot), 
			\label{theta_0_base}
		\end{equation}
		where $\{e_1(\cdot),\ldots,e_{d_n}(\cdot)\}$ is a set of B-spline basis functions and $d_n=l+m_n$ ($l$ denotes the order of the splines and $m_n$ is the number of regularly interior knots). Values $l=3$ and $m_n=3$ were considered (note that the process of optimization involved in the estimation of the MFPLSIM requires intensive computation, which forced us to select a manageable number of interior knots) and the vector of coefficients of $\theta_0$ in expression (\ref{theta_0_base}) is given by 
		\begin{equation}
			(\alpha_{01},\dots,\alpha_{0d_n})^\top=(0,1.741539,0,1.741539,-1.741539,-1.741539)^\top\label{theta0}
		\end{equation}
		(note that (\ref{theta0}) was obtained by calibrating the vector $(0,1,0,1,-1,-1)^\top$ to ensure identifiability; for details, see \citealt*{novo_2019}). The right panel in Figure \ref{fig1} shows the graphical representation of $\theta_0$.
		\item  The inner product and the link function were $\left<f,g\right>=\int_{0}^1 f(t)g(t)dt$ and $m(\left<\theta_0,\chi\right>)=\left<\theta_0,\chi\right>^3$, respectively.
		\item The i.i.d. random errors, $\varepsilon_i$ ($i=1,\ldots,n+100$), were simulated from a normal  distribution with zero mean and standard deviation equals to  $0.1$ times the standard deviation of the regression function $\sum_{j=1}^{p_n}\beta_{0j}\zeta_i(t_j)+m\left(\left<\theta_0,\mathcal{X}_i\right>\right)$.
	\end{itemize}
	
	A total of $M=100$ independent samples (i.e. $M=100$ independent copies of $\mathcal{D}$) were generated from model (\ref{mod_sim}).
	Each set $\mathcal{D}$ was split into two samples: a training sample \begin{equation}{\cal{D}}_{train}=\{(\zeta_i,\mathcal{X}_i,Y_i),\quad i=1,\dots,n\},\label{d_train}\end{equation} and a testing sample, \begin{equation}{\cal{D}}_{test}=\{(\zeta_i,\mathcal{X}_i,Y_i),\quad i=n+1, \dots,n+100 \}.\label{d_test}\end{equation}
	The training sample was used to estimate all parameters involved in (\ref{mod_sim}), while the testing sample was used to measure the quality of the corresponding predictions (i.e., the performance of the procedures) through the mean square error of prediction (MSEP):
	\begin{equation}
		\label{MSEP-sim}
		MSEP_n=\frac{1}{n_{test}}\sum_{i=n+1}^{n+n_{test}}(Y_i - \widehat{Y}_i)^2,
	\end{equation}
	where $\widehat{Y}_i$ denotes the predicted value for $Y_i$ and in this case, $n_{test}=100$. 
	For each sample, the FASSMR and the PLS procedures were applied.
	
	\subsubsection{Practical considerations} \label{T_C}
	
	In practice, various parameters have to be selected for performing the estimation associated to the FASSMR. We have the same problems if the standard PLS method is applied,
	except for selecting the splitting parameter $w=w_n$, which is specific to our new algorithm. Other parameters that should be carefully selected include the bandwidth  $h$  involved in semiparametric estimation and the tuning penalisation parameter $\lambda_k$ used in the variable selection procedure. 	For parameter $\lambda_k$, to reduce  the quantity of tuning parameters to be selected for each sample, we consider only penalty parameters of the specific form $\lambda_k= \lambda \widehat{\sigma}_{\beta_{0k,OLS}}$ with $k=1,\dots,w$, where $\beta_{0k,OLS}$ denotes the ordinary least squares (OLS) estimation of $\beta_{0k}$  in the reduced model associated to (\ref{mod_sim}) for each $w$,  and $\widehat{\sigma}_{\beta_{0k,OLS}}$ is the estimated standard deviation.

First, Epanechnikov’s kernel $K$ was used in this study owing to its low impact on
estimates. Moreover, because we want to reduce the computational time, parameters $h$, $\lambda$ and $w$ were selected via the BIC procedure. Specifically, the BIC value corresponding to $(\widehat{\pmb{\beta}}^{\pmb{1}}_{0,h,\lambda,w},\widehat{{\theta}}^{\pmb{1}}_{0,h,\lambda,w})$ (the estimate of parameter $(\pmb{\beta}^{\pmb{1}}_0,\theta^{\pmb{1}}_0)$ in the linear model (\ref{lin_mod}) obtained by minimising the profile least-squares function (\ref{func_min1})) was computed from the routine {\tt select} of the \textsf{R} package \textsf{grpreg}. We used this selector owing to its low computational cost in
comparison to cross-validation-based selectors (which are time consuming procedures).

\begin{figure}[htp]
		\centering
		\makebox{	\includegraphics[width=0.33\textwidth]{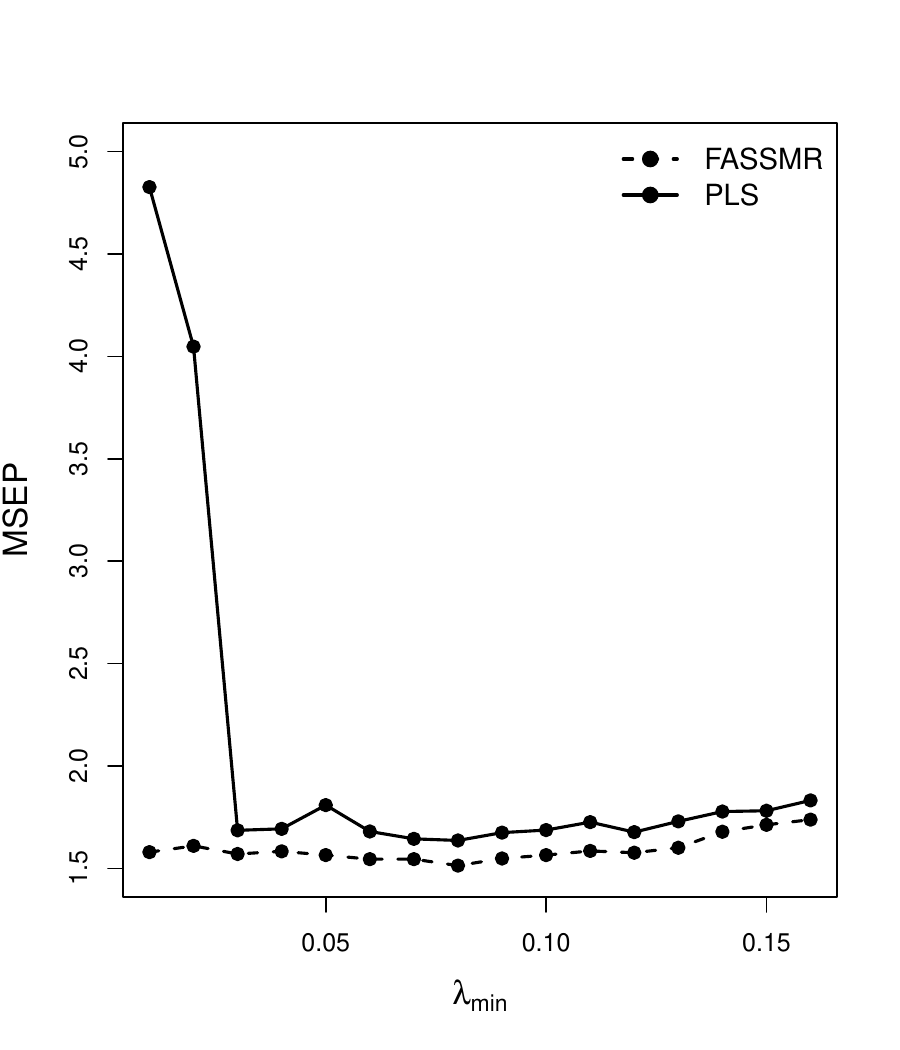}
			\includegraphics[width=0.33\textwidth]{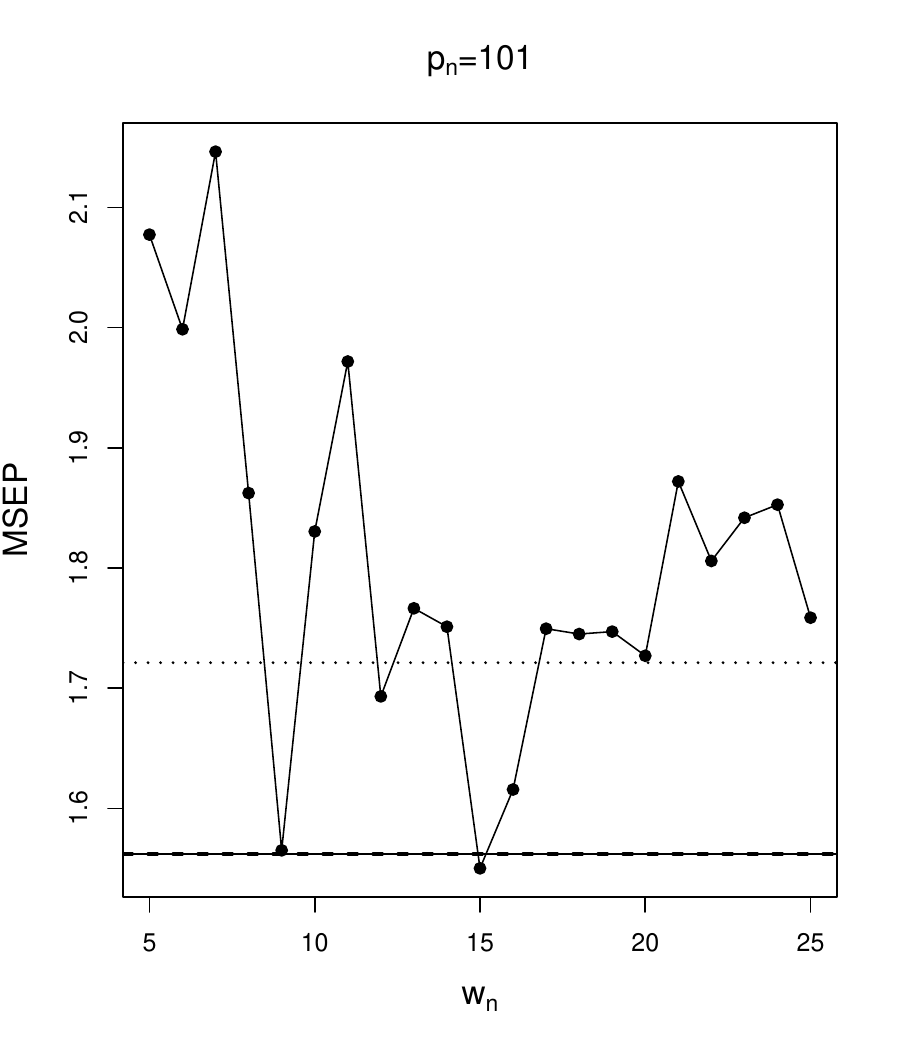}
			\includegraphics[width=0.33\textwidth]{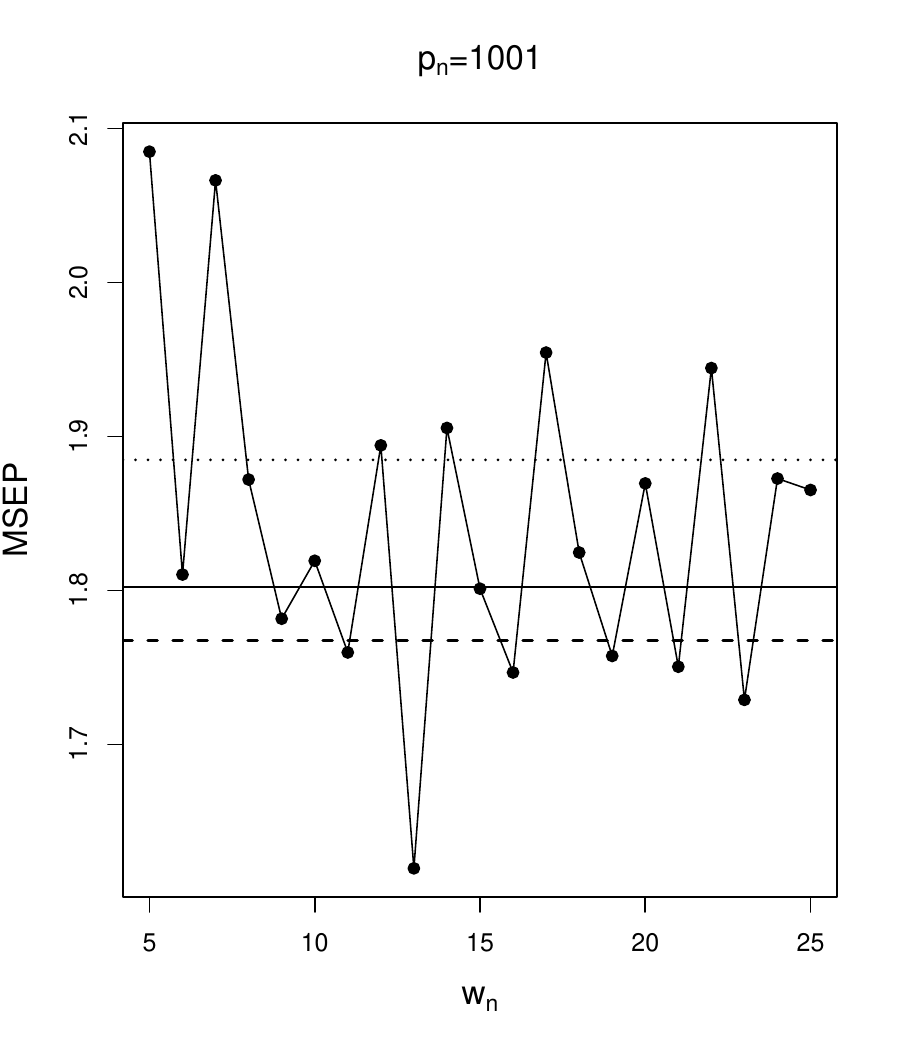}}
		\caption{ \label{fig2} Left panel: mean square error of prediction (MSEP) when the used tuning parameter $\lambda$ ($\widehat{\lambda}(\lambda_{min})$) is selected by minimising the BIC over a grid starting in $\lambda_{min}$ (considering that $\theta_0$ is known). Middle panel and right panel: 
			Mean of MSEP for each value of $w_n\in W$ for $M=10$ samples of size $n=100$ of (\ref{mod_sim}); Middle panel: case $p_n=101$; Right panel: case $p_n=1001$. In both cases, $\theta_0$ was considered to be known.  Solid horizontal lines denote the mean of MSEP for the optimal value of $w_n\in W$  selected by the BIC. Dashed line corresponds with the mean of MSEP obtained for the optimal value of $w_n\in W^*$ selected by the BIC. Dotted line is the mean of the MSEP obtained through the PLS method for the same $M=10$ samples. 
		}	
	\end{figure}

	Focusing on the penalty parameter $\lambda$, it is usually searched in a grid, $\{\lambda_{min},\ldots\}$, where $\lambda_{min}$ is the minimum value. Further, sensitivity analysis was performed on FASSMR (and the PLS) for  $\lambda_{min}$. For each value $\lambda_{min}$ considered, a grid of $100$ values, $\{\lambda_{min},\ldots\}$, was provided to the program. Then, $\widehat{\lambda}(\lambda_{min})$ was selected in this grid using the BIC and the corresponding MSEP($\widehat{\lambda}(\lambda_{min})$) was computed. The left panel in Figure \ref{fig2} shows that FASSMR is not significantly affected by $\lambda_{min}$, while small values should be discarded for the PLS method.
	
	With respect to the splitting parameter $w$, the main task involves the selection of eligible values of $w$ before applying the BIC. Figure \ref{fig2} shows the mean of MSEP over each value of $w\in W=\{5,6,\dots, 25\}$ for $M=10$ samples of size $n=100$, using $p_n=101$ (middle panel) and $p_n=1001$ (right panel). In addition, it reports the MSEP from the FASSMR when $w$ was selected with the BIC in $W$ (see the solid horizontal line) or $W^*$ (see the dashed horizontal line), where
	\begin{equation}W^*=\{10,15,20\}\label{W^*}.\end{equation} 
	Finally, the MSEP obtained from the PLS procedure is also shown (see the dotted horizontal line). The following conclusions can be derived from Figure \ref{fig2} (middle and right panels):

	\begin{itemize}
		\item FASSMR is sensitive to $w$, especially when $w$ is small. This was expected given that the FASSMR is applied to an artificial (or reduced) model, which could be different from
		the true model for small values of $w$.
		\item FASSMR improved the results of PLS in terms of MSEP for several values of $w$ between $5$ and $25$. Considering the performance of FASSMR when $w$ was selected in
		$W$ via BIC, we can conclude that the BIC is a suitable method (the corresponding MSEP
		is reasonable (see the solid horizontal line) and evidently improves the one obtained
		with the standard PLS procedure; see the dotted horizontal line).
		\item To reduce the computational time, $w$ can be selected in $W^*$ (instead of in $W$) via the BIC.

		This is because (in addition to reduce the computational time) there is no loss in terms of MSEP when  $W^*$  is used (evident from comparing the dashed and solid horizontal lines; the results are better for $p_n=1001$ using $W^*$ in comparison to when
		$W$ was used as the set for eligible values of $w$).

	\end{itemize}
	In conclusion, from now on, the set of eligible values for $w$ will be $W^*$, and the selection will be made by means of the BIC procedure.
	
	There is another minor question which is that in many practical situations, the condition  $p_n=w_n q_n$ fails. We used the solution proposed in \cite{aneirosv_2015}, based on considering non-fixed  $q_n=q_{n,k}$ values $k=1,\dots,w_n$,  when $p_n/w_n$ is not an integer. Specifically,
	\begin{equation}
		q_{n,k}= \left\{\begin{aligned}
			&[p_n/w_n]+1&  \quad k\in\{1,\dots,p_n-w_n[p_n/w_n]\},\\
			&[p_n/w_n]&\quad k\in\{p_n-w_n[p_n/w_n]+1,\dots,w_n\}, \\
		\end{aligned}
		\right.
	\end{equation}
	where $[z]$ denotes the integer part of $z\in\mathbb{R}$.
	
	Finally, to estimate $\theta_0$, a suitable set of  eligible directions, $\Theta_n^{\pmb{1}}$, should be considered.
	Accordingly, following the procedure detailed in \cite*{novo_2019}, we consider a B-spline basis with dimension $d_n=l+m_n$, $\{e_1(\cdot),\dots,e_{d_n}(\cdot)\}$, for generating directions:
	\begin{equation}
		\theta^{\pmb{1}}(\cdot)=\sum_{j=1}^{d_n}\alpha_je_j(\cdot).
		\label{theta_base}
	\end{equation}
	In the expression (\ref{theta_base}), the vector of coefficients $(\alpha_1,\dots,\alpha_{d_n})$ was obtained as follows:
	\begin{enumerate}
		\item We considered $(\alpha_1^*,\dots,\alpha_{d_n}^*)\in\mathcal{M}^{d_n}$, where $\mathcal{M}=\{-1,0,1\}$ was used as the set of  seed-coefficients (owing to the high computational cost of the estimation process, we used
		a manageable number of seed-coefficients).
		\item $(\alpha_1^*,\dots,\alpha_{d_n}^*)$ were calibrated such $\left<\theta^{\pmb{1}},\theta^{\pmb{1}}\right>=1$ and $\theta^{\pmb{1}}(0.5)>0$.
	\end{enumerate}
	The resultant coefficients of steps 1 and 2 are the desired  $(\alpha_1,\dots,\alpha_{d_n})$. Therefore, in practice, $\Theta_n^{\pmb{1}}$ is composed by the directions generated from (\ref{theta_base}).

	\subsubsection{Results}\label{sim1-res}
	
	\begin{table}[htp]
		\caption{\label{Comp_Tim}Computational time in seconds required for estimating one sample of model (\ref{mod_sim}) using the PLS procedure and FASSMR for different values of $n$ and $p_n$. In the case of FASSMR, the eligible values for $w_n$ are those belonging to $W^*$. The results were obtained with a computer with the following features: Intel Core i7-7700HQ CPU, 8 GB RAM, 1 TB HDD, 256 GB SSD.}
		\centering
		\scalebox{0.9}{\begin{tabular}{|lr|rrrrr|}
				\hline
				$n$& Method & $p_n=101$ & $p_n=201$ & $p_n=501$ & $p_n=1001$ & $p_n=10001$\\
				\hline
				\multirow{2}{1cm}{$100$}& 	PLS & 727.55 &1324.2 & 2571.52 & 4959.70 & 43137.25\\
				&FASSMR & 374.28& 362.67& 367.23 &365.14 & 357.17 \\
				\hline
				\multirow{2}{1cm}{$200$}&	PLS & 1089.18& 2625.83& 7211.37&14823.14&153540.27 \\
				&	FASSMR & 1058.52& 1034.00& 1032.03&1008.02&702.67 \\
				\hline	
				\multirow{2}{1cm}{$300$}&	PLS &3341.82 &8091.13 & 20537.98 &33868.11& 224890.17\\
				&FASSMR & 3184.35& 3412.45 &2361.60&3123.95  & 2448.00 \\	
				\hline
		\end{tabular}}
	\end{table}
	It is evident from Table \ref{Comp_Tim} that the PLS method is inefficient for large values of $p_n$. By
	contrast, FASSMR led to reduced computational time; consequently, even for significantly
	large values of $p_n$ the results were obtained in a reasonable amount of time.  Furthermore, Table \ref{MSEP_FASSMR_PLS} demonstrates that when FASSMR was used, no loss was observed in the MSEP.
	It
	should be noted that Table \ref{MSEP_FASSMR_PLS} denotes the results obtained using $100$ samples; we did not
	consider large values of $n$ and $p_n$ owing to the significant computational time required by
	the PLS procedure for estimating one sample (see again Table \ref{Comp_Tim}).

	\begin{table}[htp]
		\centering
		\caption{ \label{MSEP_FASSMR_PLS}For $M=100$ samples of model (\ref{mod_sim}), mean of MSEP for PLS procedure and FASSMR.}
		\scalebox{0.9}{
			\begin{tabular}{lr|rr|rr|rr|}
				\cline{3-8}
				&& \multicolumn{6}{|c|}{MSEP} \\
				\cline{3-8} 
				& &\multicolumn{2}{|c|}{$p_n=101$}&\multicolumn{2}{c|}{$p_n=201$}&\multicolumn{2}{c|}{$p_n=501$}\\
				\hline
				\multicolumn{1}{|l}{$n$}& Method & Mean & SD & Mean & SD& Mean & SD\\
				
				\hline	
				\multicolumn{1}{|c}{\multirow{2}{1cm}{$100$}}&	PLS & 1.2572& 0.9546&1.3213&1.0078&1.3126&0.9688 \\
				\multicolumn{1}{|c}{}	&FASSMR  &  1.1579&1.0335 &  1.2694& 1.0675& 1.2025&0.9507\\
				\hline
				\multicolumn{1}{|c}{\multirow{2}{1cm}{$200$}}&	PLS & 0.7662 & 0.4630& 0.7617 & 0.4209 & 0.7194 & 0.4606\\
				\multicolumn{1}{|c}{}	&FASSMR  &0.6984&0.4274& 0.8049&0.5000&  0.7357&0.4860 \\
				
				\hline
		\end{tabular}}
		
	\end{table}

	\subsection{Conclusions and open questions} 
	FASSMR enabled us to obtain the variable selection and estimation of model (\ref{model}) in a
	reasonable amount of time, even for significantly large values of $p_n$. The simulation study
	indicates that the developed algorithm clearly surpasses the standard PLS procedure in
	terms of computational time without loss in predictive ability. Moreover, the application in each
	scenario of the Diebold-Mariano test in Table \ref{MSEP_FASSMR_PLS} (for comparing the forecast accuracy of the two prediction methods) provides further conclusions; in some scenarios, no significant
	difference was observed between PLS and FASSMR, and in the scenarios wherein significant
	differences were observed, FASSMR demonstrated improvement in the predictive ability.
	
	However, even though the computational time was improved, the set of relevant variables
	could not be exactly obtained in many cases. As explained in the Appendix, this can be
	confirmed by the asymptotic analysis derived from Proposition \ref{proposition}.This may be inconvenient
	in some real data applications, wherein it is crucial to know the influential variables. However,
	situations of grouped impact points (GIPs) are common, wherein the FASSMR cannot provide
	the full set of influential variables. Thus, one might wonder if an additional algorithm,
	based on the functional origin of scalar linear covariates, that can select a more precise
	set of impact points without destroying the main features of the FASSMR, particularly its fast
	implementation, can be developed.
	
	\section{IASSMR: a refined variable selection algorithm }
	As discussed above, the FASSMR should refined to obtain a more precise set of selected
	impact points. The concept of an improved algorithm for sparse semiparametric regression
	(IASSMR) involves adding a second stage that considers $q_n$ variables in the neighbourhood of those selected in the first stage by the FASSMR. Then, a second variable selection procedure
	is applied among this new set of variables. 
	
	\subsection{Procedure}\label{proce}
	For developing the IASSMR, sample (\ref{sample}) is split into two independent subsamples asymptotically of the same size $n_1\sim n_2\sim n/2$; one of them to be used in the first stage of the method and the other in the second stage:
	\begin{equation*}
		\mathcal{E}^{\pmb{1}}=\{(\zeta_i,\mathcal{X}_i,Y_i),\quad i=1,\dots,n_1\},
	\end{equation*} 
	\begin{equation*}
		\mathcal{E}^{\pmb{2}}=\{(\zeta_i,\mathcal{X}_i,Y_i),\quad i=n_1+1,\dots,n_1+n_2=n\}.
	\end{equation*} 
	Henceforth, the superscript $\pmb{s}$ with $\pmb{s}=\pmb{1},\pmb{2}$ indicates the stage of the method in which the sample, function, variable or parameter is involved.

	\begin{description}
		\item[First stage.] The FASSMR is applied using only subsample $\mathcal{E}^{\pmb{1}}$:
		\begin{enumerate}
			\item The procedure is started the variables belonging to 
			$\mathcal{R}_n^{\pmb{1}}$, see (\ref{initial_set}). The MFPLSIM  is transformed into a linear model  as in (\ref{lin_mod}). Note that because we only use $\mathcal{E}^{\pmb{1}}$,
			we obtain an $n_1\times n_1$-matrix of local weights 
			$\pmb{W}_{h,\theta}^{\pmb{1}}=\left(w_{n_1,h,\theta}(\mathcal{X}_i,\mathcal{X}_{\ell})\right)_{i,\ell=1,\dots,n_1},$ where $w_{n_1,h,\theta}(\mathcal{X}_i,\mathcal{X}_{\ell})$ has been defined in $(\ref{local_weights})$. Now, for each $\theta\in\mathcal{H}$, we denote
			$\widetilde{\pmb{Y}}_{\theta}^{\pmb{1}}=\left(\pmb{I}-\pmb{W}_{h,\theta}^{\pmb{1}}\right)\pmb{Y}^{\pmb{1}}$ with $\pmb{Y}^{\pmb{1}}=(Y_1,\dots,Y_{n_1})^{\top}$ and $\widetilde{\pmb{\zeta}}_{\theta}^{\pmb{1}}=\left(\pmb{I}-\pmb{W}_{h,\theta}^{\pmb{1}}\right)\pmb{\zeta}^{\pmb{1}},$
			where, committing an abuse of notation, we denote the $n_1\times w_n$ matrix  $(\zeta_i(t_k^{\pmb{1}}), 1\leq i\leq n_1, 1\leq k\leq w_n)$ by $\pmb{\zeta}^{\pmb{1}}$.
			\item
			The standard PLS variable selection procedure is applied in the set $\mathcal{R}_n^{\pmb{1}}$ by minimising the PLS criterion over the pair $(\pmb{\beta}^{\pmb{1}},\theta^{\pmb{1}})$ with $\pmb{\beta}^{\pmb{1}}\in\mathbb{R}^{w_n}$ and  $\theta^{\pmb{1}}\in \Theta_n^{\pmb{1}}$:
			\begin{equation}
				\mathcal{Q}^{\pmb{1}}\left(\pmb{\beta}^{\pmb{1}},\theta^{\pmb{1}}\right)=\frac{1}{2}\left(\widetilde{\pmb{Y}}_{\theta^{\pmb{1}}}^{\pmb{1}}-\widetilde{\pmb{\zeta}}_{\theta^{\pmb{1}}}^{\pmb{1}}\pmb{\beta}^{\pmb{1}}\right)^{\top}\left(\widetilde{\pmb{Y}}_{\theta^{\pmb{1}}}^{\pmb{1}}-\widetilde{\pmb{\zeta}}_{\theta^{\pmb{1}}}^{\pmb{1}}\pmb{\beta}^{\pmb{1}}\right)+n_1\sum_{k=1}^{w_n}\mathcal{P}_{\lambda_{k_n}}\left(|\beta_{k}^{\pmb{1}}|\right).
				\label{func_min12}
			\end{equation}
			
			\item We obtain $(\widehat{\pmb{\beta}}_0^{\pmb{1}},\widehat{\theta}_0^{\pmb{1}})$ by minimising (\ref{func_min12}); then, $\zeta(t_k^{\pmb{1}})$ is selected in $\mathcal{R}_n^{\pmb{1}}$ if, and only if, $\widehat{\beta}_{0k}^{\pmb{1}}\not =0$.
		\end{enumerate}
		\item[Second stage.]  The variables in the neighborhood of those selected in the first stage are
		included; then, the PLS procedure is performed again. Here, we only consider subsample $\mathcal{E}^{\pmb{2}}$. Specifically:
		\begin{enumerate}
			\item A new set of variables is considered:
			\begin{equation*}
				\mathcal{R}_n^{\pmb{2}}=\bigcup_{\left\{k,\widehat{\beta}_{0k}^{\pmb{1}}\not=0\right\}}\left\{\zeta(t_{(k-1)q_n+1}),\dots,\zeta(t_{kq_n})\right\}.
			\end{equation*}
			Here, $r_n=\sharp(\mathcal{R}_n^{\pmb{2}})$; thus, we can rename the variables in $\mathcal{R}_n^{\pmb{2}}$ as follows:
			\begin{equation*}
				\mathcal{R}_n^{\pmb{2}}=\left\{\zeta(t_1^{\pmb{2}}),\dots,\zeta(t_{r_n}^{\pmb{2}})\right\},
			\end{equation*}
			and consider  the following model
			\begin{equation}
				Y_i=\sum_{k=1}^{r_n}\beta_{0k}^{\pmb{2}}\zeta_i(t_k^{\pmb{2}})+m^{\pmb{2}}\left(\left<\theta_0^{\pmb{2}},\mathcal{X}_i\right>\right)+\varepsilon_i^{\pmb{2}}.
				\label{mod_r2}
			\end{equation}
			
			\item Similar to the first stage, model (\ref{mod_r2}) is transformed into a linear model analogously to in (\ref{lin_mod}):
			\begin{equation}
				Y_i-\mathrm{E}\left(Y_i|\left<\theta_0^{\pmb{2}},\mathcal{X}_i\right>\right)=\sum_{k=1}^{r_n}\beta_{0k}^{\pmb{2}}\left(\zeta_i(t_k^{\pmb{2}})-\mathrm{E}\left(\zeta_i(t_k^{\pmb{2}})|\left<\theta_0^{\pmb{2}},\mathcal{X}_i\right>\right)\right)+\varepsilon_i^{\pmb{2}},
				\label{lin_mod2}
			\end{equation}
		however, now we use the subsample $\mathcal{E}^{\pmb{2}}$ to obtain the estimate of the conditional expectations.  Therefore, $\pmb{W}_{h,\theta}^{\pmb{2}}$ is obtained analogously to $\pmb{W}_{h,\theta}^{\pmb{1}}$ but employing  $\mathcal{E}^{\pmb{2}}$ instead of $\mathcal{E}^{\pmb{1}}$.
			Similar to the previous stage, for each $\theta\in\mathcal{H}$, 
			$\widetilde{\pmb{Y}}_{\theta}^{\pmb{2}}=\left(\pmb{I}-\pmb{W}_{h,\theta}^{\pmb{2}}\right)\pmb{Y}^{\pmb{2}}$ with $\pmb{Y}^{\pmb{2}}=(Y_{n_1+1},\dots,Y_{n})^{\top}$ and  $\widetilde{\pmb{\zeta}}_{\theta}^{\pmb{2}}=\left(\pmb{I}-\pmb{W}_{h,\theta}^{\pmb{2}}\right)\pmb{\zeta}^{\pmb{2}}$
			with  $\pmb{\zeta}^{\pmb{2}}$ the $n_2\times r_n$ matrix $\left(\zeta_{i}(t_k^{\pmb{2}}), n_1+1\leq i\leq n, 1\leq k\leq r_n\right)$,  and $\pmb{\beta}^{\pmb{2}}=(\beta_1^{\pmb{2}},\dots,\beta_{r_n}^{\pmb{2}})^{\top}$.

			\item The PLS procedure is applied in set $\mathcal{R}_n^{\pmb{2}}$ by minimising the profile least squares function over the pair $(\pmb{\beta}^{\pmb{2}},\theta^{\pmb{2}})$ with $\pmb{\beta}^{\pmb{2}}\in\mathbb{R}^{r_n}$ and  $\theta^{\pmb{2}}\in \Theta_n^{\pmb{2}}\subset\mathcal{H}$:
			\begin{equation}
				\mathcal{Q}^{\pmb{2}}\left(\pmb{\beta}^{\pmb{2}},\theta^{\pmb{2}}\right)=\frac{1}{2}\left(\widetilde{\pmb{Y}}_{\theta^{\pmb{2}}}-\widetilde{\pmb{\zeta}}_{\theta^{\pmb{2}}}^{\pmb{2}}\pmb{\beta}^{\pmb{2}}\right)^{\top}\left(\widetilde{\pmb{Y}}_{\theta^{\pmb{2}}}-\widetilde{\pmb{\zeta}}_{\theta^{\pmb{2}}}^{\pmb{2}}\pmb{\beta}^{\pmb{2}}\right)+n_2\sum_{k=1}^{r_n}\mathcal{P}_{\lambda_{k_n}}\left(|\beta_{k}^{\pmb{2}}|\right).
				\label{func_min2}
			\end{equation}
			\item The minimiser of criterion $\mathcal{Q}^{\pmb{2}}(\cdot,\cdot)$ is denoted by $\left(\widehat{\pmb{\beta}}_0^{\pmb{2}},\widehat{\theta}_0^{\pmb{2}}\right)$. At the end of the second stage, $\zeta(t_k^{\pmb{2}})$ in $\mathcal{R}_n^{\pmb{2}}$ and the associated coefficient, $\widehat{\beta}_{0k}^{\pmb{2}}$ is non-null.
		\end{enumerate}
	\end{description}
	
	\begin{remark} The theoretical considerations involving subsets $\Theta_n^{\pmb{2}}$ and $\Theta_n^{\pmb{1}}$ and the local-minimiser in the IASSMR are the same as those given in Remark \ref{remark_t} for the FASSMR.  
	\end{remark}
	\subsection{Outputs of IASSMR}
	At the end of this two-stage procedure, a variable $\zeta(t_j)\in\{\zeta(t_1),\dots,\zeta(t_{p_n})\}$ is selected if, and only if, it belongs to $\mathcal{R}_n^{\pmb{2}}$ and its estimated coefficient in the second stage, $\widehat{\beta}_{0k_j}^{\pmb{2}}$,  is non-null. Therefore, the following estimated set of relevant variables is obtained:
	\begin{equation}
		\widehat{S}_n=\left\{j=1,\dots,p_n,\  \textrm{such that } t_j=t_{k_j}^{\pmb{2}}, \textrm{ with } \zeta(t_{k_j}^{\pmb{2}})\in\mathcal{R}_n^{\pmb{2}} \textrm{ and } \widehat{\beta}_{0k_j}^{\pmb{2}}\not=0\right\}.\label{S_n_est}
	\end{equation}
	In this case, an estimator for the linear coefficients and direction $\theta_0$ can be naturally obtained
	using the estimations from the second stage of the algorithm, i.e.,
	\begin{eqnarray}
		\widehat{\beta}_{0j}&=&\left\{ 
		\begin{aligned} \widehat{\beta}_{0k_j}^{\pmb{2}}\quad  \textrm{ if } j\in \widehat{S}_n,\\
			0 \quad \textrm{ otherwise,} \label{est_beta0} \\
		\end{aligned}\right.
	\end{eqnarray}
	
	\begin{eqnarray}
		\widehat{\theta}_0&=&\widehat{\theta}^{\pmb{2}}_0.     \label{est_theta0}            
	\end{eqnarray}
	
	Denoting by $\widehat{\pmb{\beta}}_0$ the vector of estimated linear coefficients, an estimator for the function $m_{\theta_0}(\cdot)\equiv m(\left<\theta_0,\chi\right>)$ can be obtained by smoothing the residuals of the linear component, similar to (\ref{est_m}), but using $\widehat{\beta}_{0j}$ and $\widehat{\theta}_0$ obtained in (\ref{est_beta0}) and (\ref{est_theta0}), respectively. In other words, $\widehat{m}_{\widehat{\theta}_0}(\chi)=\widehat{m}_{\widehat{\theta}_0^{\pmb{2}}}^{\pmb{2}}(\chi)$.

	\begin{remark} Now, we present some remarks regarding the considered subsamples ($\mathcal{E}^{\pmb{1}}$ and $\mathcal{E}^{\pmb{2}}$ for first and second stages, respectively) and the theoretical complexity of the algorithm. Focusing on $\mathcal{E}^{\pmb{1}}$ and $\mathcal{E}^{\pmb{2}}$, it should be noted that because they are different, and therefore,
		independent, the bias of selection is avoided, thus facilitating the proof of our asymptotic
		results. In addition, although we selected $n_1\sim cn$ and $n_2\sim n-n_1$ for $c=1/2$ (maybe the natural choice), one could consider any value $0<c<1$ (this does not affect the asymptotic properties; however, in some scenarios, such as where the sample size ($n$) is too small, it could be convenient to consider $c\neq 1/2$). To obtain the theoretical complexity of
		the algorithm, the following should be considered: (i) the construction of linear model in
		the first stage, (ii) the application of the variable selection procedure to this linear model,
		(iii) the construction of the linear model in the second stage, and (iv) the application of the
		variable selection procedure to this linear model. For a fixed value $\theta \in \Theta_n$ (we considered $\Theta_n=\Theta_n^{\pmb{1}}=\Theta_n^{\pmb{2}}$) and given tuning parameters $h$, $w_n$ and $\lambda$, the theoretical complexities related to the first stage ((i) and (ii)) are $O(n^2w_n)$ and $O(nw_n)$ respectively. For the second stage ((iii) and (iv)), the theoretical complexities depend on $r_n$ (the number of covariates in the linear model corresponding to the second stage), which is a random variable. It is evident
		in the study by \cite{aneirosv_2014} that under suitable conditions, it verifies $r_n=O(s_n)$ with probability 1 (w.p.1). Therefore, we the theoretical complexities related to (iii) and (iv) are $O(n^2s_n)$ and $O(ns_n)$ w.p.1, respectively. In summary, the theoretical complexity of the proposed IASSMR is $O(n^2w_n\sharp\Theta_n)+O(n^2s_n\sharp\Theta_n)$ w.p.1. Therefore, in the usual case where $\max\{s_n,w_n\} \ll p_n$, the IASSMR algorithm is expected to be significantly faster
		than the standard one (this will be more evident in situations where $\max\{s_n,w_n\}\ll n \ll p_n$) but slower than the FASSMR algorithm (especially in situations where $s_n\gg w_n$); for the theoretical complexities of the IASSMR algorithm and the standard method, see Remark \ref{remark_tt}.
		
		\label{remark_ttt}
	\end{remark}

	\subsection{A simulation study}
	
	In this section, we discuss the Monte Carlo studies performed to compare the finite
	sample behavior of FASSMR and IASSMR in two different frameworks, i.e., a first scenario with spaced impact points and a  second one with grouped impact points (GIPs).
	\subsubsection{First design: spaced impact points}\label{des1}
	
\begin{figure}[htp]
		\centering
		\includegraphics[width=0.33\textwidth]{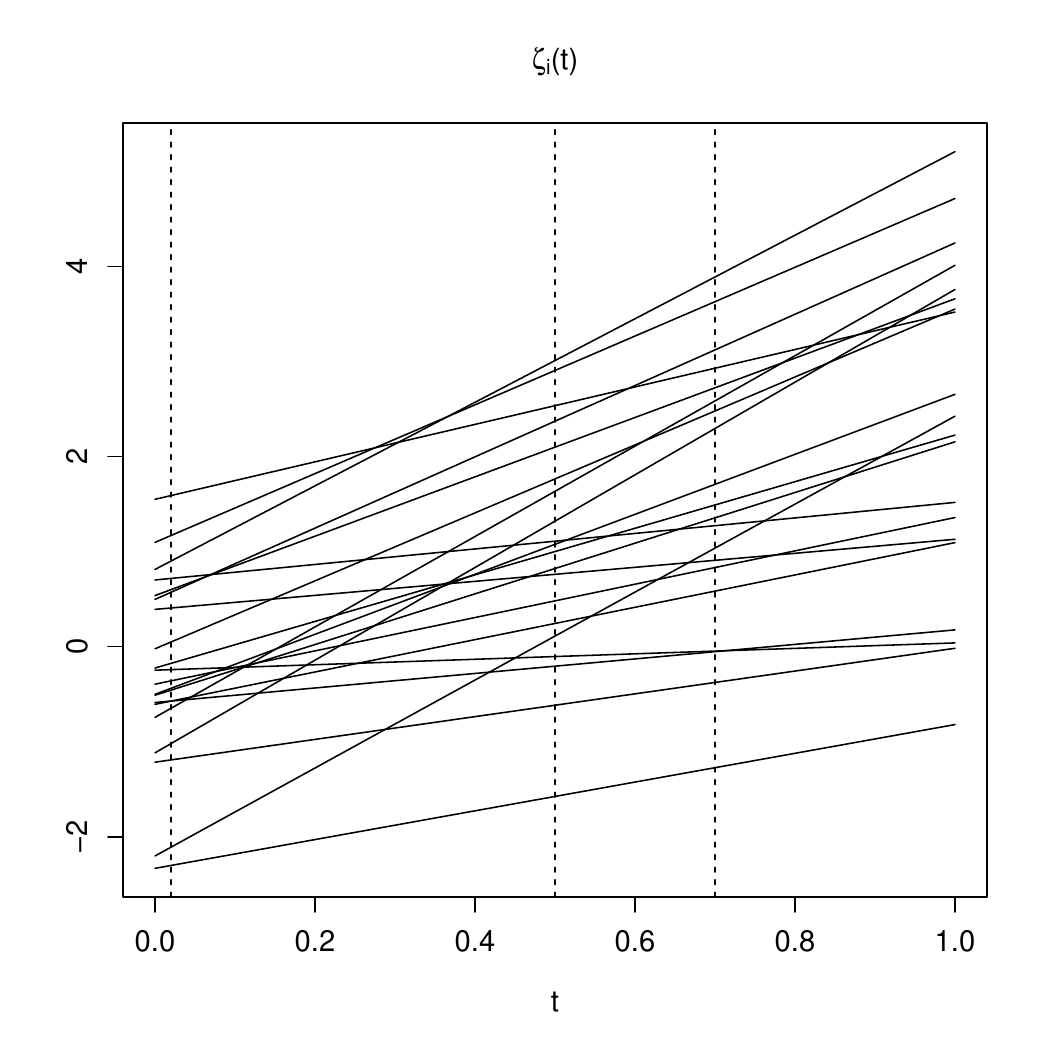}
		\caption{Sample of $20$ lines generated from (\ref{parte_lineal}), along with impact points (dotted vertical lines) at instants $t_{j1}=0.02$, $t_{j2}=0.50$  and $t_{j3}=0.70$.}
		\label{rectas}
	\end{figure}

	Observations i.i.d.  $\mathcal{D}=\{(\zeta_i,\mathcal{X}_i,Y_i),\quad i=1,\dots,n+100\}$ were generated from the model
	\begin{equation}
		Y_i=\sum_{j=1}^{p_n}\beta_{0j}\zeta_i(t_j)+m\left(\left<\theta_0,\mathcal{X}_i\right>\right)+\varepsilon_i. \label{mod_sim1}
	\end{equation}
	Here,
	\begin{itemize}
		\item The curves involved in the non-linear part, $\mathcal{X}_i=\mathcal{X}_{a_i,b_i,c_i}$ were generated from expression (\ref{X-sim}); 
		however, the random variables $a_i$, $b_i$ and $c_i$ (independent between and within vectors $(a_i,b_i,c_i)^{\top}$) are uniformly distributed on the interval $[0,5]$. These curves were discretised on the same grid of $100$ equispaced points in $[0,1]$.
		\item  ${t_j}$ denotes equispaced points in $[0,1]$, with $t_1=0$ and $t_{p_n}=1$.
		\item The curves involved in the linear component were generated from the expression
		\begin{equation}
			\zeta_i(t_j)=c_it_j+d_i \label{parte_lineal}
		\end{equation}
		where $d_i$ is normally distributed with mean and standard deviation equals of $0$ and $1$, respectively, and $c_i$ was defined in the first item. Consequently, there exists some dependence between $\mathcal{X}$ and $\zeta$. In addition, we considered only three non-null coefficients:
		$\beta_{0j_1}=4$, $\beta_{0j_2}=3$ and $\beta_{0j_3}=-3.2$, with impact points $t_{j1}=0.02$, $t_{j2}=0.50$  and $t_{j3}=0.70$ (left panel in Figure \ref{rectas} shows $100$ curves $\zeta_i$  with influential points marked in dotted vertical lines).	
		\item The true direction of projection ($\theta_0$),  inner product ($\left<\cdot,\cdot\right>$), link function ($m(\cdot)$) and random errors ($\varepsilon_i$) were generated as shown in Section \ref{des}.
	\end{itemize}

	A total of $M=100$ independent samples were generated from  (\ref{mod_sim1}), which were divided in $\mathcal{D}_{train}$ (see (\ref{d_train})) and $\mathcal{D}_{test}$ (see (\ref{d_test})). Values $p_n\in\{101,201,501,1001,10001\}$ were considered. In this case, instead of fixing the sample size to be equal for the two methods, we are going to fix the sample size of the first step (the only step in the FASSMR), $n_1=100$, and varied the sample size of the second step, $n_2$ (we considered $n_2\in\{0,100,200\}$; case $n_2=0$ corresponds with the FASSMR). To perform estimation using each method, we followed the technical considerations specified in Section \ref{T_C}. It should be noted that, similar to FASSMR, we used $W^*$ (see (\ref{W^*})) as set of eligible values of $w_n$ in IASSMR and the set of eligible directions, $\Theta_n^{\pmb{2}}$, was generated in the same way as $\Theta_n^{\pmb{1}}$ (see (\ref{theta_base})).

In this simulation study, we attempted to compare the practical behavior of IASSMR and
FASSMR with respect to the computational efficiency, MSEP (\ref{MSEP-sim}), and precision of the impact point selection. In other words, we attempted to quantify the accuracy of set $\widehat{S}_n$ obtained using each procedure. However, the continuous origin of linear covariates makes it difficult to difference the effect of points that are considerably close in the discretisation. %and this fact is even more relevant in the GIP-MFPLSIM case.
	Therefore, comparing $\widehat{S}_n$ with $S_n$ (via classical measures as FDR, specificity and sensibility) can be inappropriate, but the following sets can be considered:
	\begin{eqnarray}
		I_{n}=[0.00,0.05]\cup[0.47,0.53]\cup[0.67,0.73],\nonumber\\
		\overline{I}_{n}=(0.05,0.47)\cup(0.53,0.67)\cup(0.73,1].\nonumber
	\end{eqnarray}
	and to classify as well chosen all those  selected points belonging to $I_{n}$ and as wrongly chosen those belonging to $\overline{I}_{n}$. In other words, denoting the set of true impact points as 
	\begin{equation}T_n=\{t_j, \ j=1,\dots,p_n, \ \beta_{0j}\not=0\},\label{T_n}\end{equation} and its estimation as \begin{equation*}\widehat{T}_n=\{t_j, \ j=1,\dots,p_n,\ \widehat{\beta}_{0j}\not=0\},\label{hat_T_n}\end{equation*}
	we quantified Right$=\sharp\{I_{n}\cap \widehat{T}_{n}\}$ and Wrong$=\sharp\{\overline{I}_{n}\cap \widehat{T}_{n}\}$ for IASSMR and FASSMR.

	\subsubsection{Practical considerations} \label{T_C2}
	
Similar to FASSMR, various tuning parameters must be selected for performing the
estimations associated with IASSMR. Here, we focus on the selection of parameters $h$, $\lambda$ and $w$ given that two stages were considered in IASSMR, some clarifications are
required regarding the BIC procedure to be used. The goal is to select these parameters
such that the final estimator in the second stage (equivalently, in the second model M2 (\ref{lin_mod2})) 
	%(ie, the estimator $(\widehat{\pmb{\beta}}^{\pmb{2}}_{0,h,\lambda,w},\widehat{{\theta}}^{\pmb{2}}_{0,h,\lambda,w})$ of the parameter $(\pmb{\beta}^{\pmb{2}}_0,\theta^{\pmb{2}}_0)$ in the linear model (\ref{lin_mod2}) obtained by minimizing the profile least-squares function (\ref{func_min2})) 
	achieves the minimum value for the BIC. Specifically, because the covariates in M2 depend on
	the covariates selected in the first model M1 (the model in stage 1), we first select, for each $w$, the covariates in M1 using the BIC procedure to select the corresponding parameters $h_w^{\pmb{1}}$ and $\lambda_w^{\pmb{1}}$. Then, after constructing M2$=$M2$_w$, the BIC procedure is applied again to select the parameters $h_w^{\pmb{2}}$ and $\lambda_w^{\pmb{2}}$  corresponding to the estimators of $(\pmb{\beta}^{\pmb{2}}_0,\theta^{\pmb{2}}_0)$ in M2$_w$. Finally, if BIC$_w^{\pmb{2}}=$ BIC$(h_w^{\pmb{2}},\lambda_w^{\pmb{2}})$ (the BIC value corresponding to such estimators), the selected parameters are $w^{\pmb{2}}_{opt}$, $h^{\pmb{2}}_{w^{\pmb{2}}_{opt}}$ and $\lambda^{\pmb{2}}_{w^{\pmb{2}}_{opt}}$, where $w^{\pmb{2}}_{opt}=\arg \min$ BIC$_w^{\pmb{2}}$.

	\subsubsection{Results}
	
	Tables \ref{TIME_FASSMR_IASSMR}--\ref{IP_FASSMR_IASSMR_1} summarise the effect of adding a second step to the FASSMR in terms of computational efficiency, MSEP and precision in impact point selection, respectively, as well as the influence of the sample size in this second step. As expected, the second stage increases the total time required for estimation and this increase is larger the bigger is the size of the discretisation (owing to the construction of  $\mathcal{R}_n^{\pmb{2}}$)%, but it is still considerably lower than the estimation time needed by the PLS method (see Table \ref{Comp_Tim})
	. However,  from  Tables \ref{MSEP_FASSMR_IASSMR} and \ref{IP_FASSMR_IASSMR_1} it is evident that the MSEP and precision of impact point selection improved owing to the
	second stage. Note that we did not consider large values for $p_n$ due to the significant
	computational time required for estimating one sample in the case of IASSMR. Furthermore,
	by analysing the effect of increase in $p_n$ in Table \ref{MSEP_FASSMR_IASSMR},  it is evident that IASSMR is less affected
	than FASSMR; in the case of IASSMR, the results did not deteriorate.

	\begin{table}[htp]
		\centering
		\caption{ \label{TIME_FASSMR_IASSMR}For $M=100$ samples from MFPLSIM (\ref{mod_sim1}), time in seconds required to make the estimations for one sample of size $n=n_1+n_2$.  The results were obtained with a computer with the following features: Intel Core i7-7700HQ CPU, 8 GB RAM, 1 TB HDD, 256 GB SSD.}
		\scalebox{0.9}{
			\begin{tabular}{lr|rrrrr|}
				
				\hline 
				\multicolumn{1}{|l}{$n_2$}& Method &\multicolumn{1}{|c}{$p_n=101$}&\multicolumn{1}{c}{$p_n=201$}&\multicolumn{1}{c}{$p_n=501$}&\multicolumn{1}{c}{$p_n=1001$}&\multicolumn{1}{c|}{$p_n=10001$}\\
				\hline	
				\multicolumn{1}{|c}{$0$}&	FASSMR & 203.00 & 202.75& 202.64 & 203.20& 210.21\\
				\multicolumn{1}{|c}{$100$}	&IASSMR  &486.94 & 678.60 & 1243.92 &2332.59&21278.42\\
				
				\multicolumn{1}{|c}{$200$}&		IASSMR & 840.93& 1153.92& 2435.96 & 6636.79&43554.89\\
				\hline
		\end{tabular}}
		
	\end{table}

	\begin{table}[htp]
		\centering
		\caption{ \label{MSEP_FASSMR_IASSMR}For $M=100$ samples from MFPLSIM (\ref{mod_sim1}), mean of MSEP for  FASSMR and IASSMR using $n_1=100$ ($n=n_1+n_2$).}
		\scalebox{0.9}{
			\begin{tabular}{lr|rr|rr|rr|}
				\cline{3-8}
				&& \multicolumn{6}{|c|}{MSEP} \\
				\cline{3-8} 
				& &\multicolumn{2}{|c|}{$p_n=101$}&\multicolumn{2}{c|}{$p_n=201$}&\multicolumn{2}{c|}{$p_n=501$}\\
				\hline
				\multicolumn{1}{|l}{$n_2$}& Method & Mean & SD & Mean & SD& Mean & SD\\
				
				\hline	
				\multicolumn{1}{|c}{$0$}&	FASSMR &0.5877 & 0.3666&0.6122&0.4498&0.6164& 0.4636\\
				\multicolumn{1}{|c}{$100$}	&IASSMR  &0.3888  &0.1563 &0.3965  &0.1689 &0.3800 &0.1543\\
				
				\multicolumn{1}{|c}{$200$}&		IASSMR & 0.3174 & 0.1049& 0.3161 & 0.1045& 0.3166 &0.1040\\
				\hline
		\end{tabular}}
		
	\end{table}

	\begin{table}[htp]
		\centering
		\caption{\label{IP_FASSMR_IASSMR_1} For $M=100$ samples from MFPLSIM (\ref{mod_sim1}), the mean  number of variables correctly selected ($\sharp\{I_{n}\cap \widehat{T}_{n}\}$)  and wrongly selected ($\sharp\{\overline{I}_{n}\cap \widehat{T}_{n}\}$) by FASSMR and IASSMR using $n_1=100$ ($n=n_1+n_2$).}
		\scalebox{0.9}{
			\begin{tabular}{cc|cc|cc|cc|}
				\cline{3-8}
				&&\multicolumn{2}{c|}{$p_n=101$}&\multicolumn{2}{c|}{$p_n=201$}&\multicolumn{2}{c|}{$p_n=501$}\\
				\hline
				\multicolumn{1}{|l}{$n_2$}& Method& Right & Wrong & Right & Wrong & Right & Wrong\\
				\hline	
				\multicolumn{1}{|c}{$0$}&	FASSMR &1.24 & 1.60&1.28
				&1.36&1.29 &1.50\\
				\multicolumn{1}{|c}{$100$}& IASSMR &1.31 & 1.21 &1.27  &1.15 &1.34 &0.90\\
				\multicolumn{1}{|c}{$200$}&	IASSMR &1.36 &1.16  & 1.30&1.18 &1.33
				& 0.88
				\\
				\hline
		\end{tabular}}
		
	\end{table}
	
	\subsubsection{Second design: grouped impact points} \label{des2}
	Similar to Section \ref{S1},  observations i.i.d.  $\mathcal{D}=\{(\zeta_i,\mathcal{X}_i,Y_i),\quad i=1,\dots,n+100\}$ were generated using $n\in\{100,200,300\}$ and $p_n\in\{101,201,501,1001,10001\}$; however, considering
	the following modification of model (\ref{mod_sim}):
	\begin{equation}
		Y_i=\sum_{j=1}^{p_n}\beta_{0j}\zeta_i(t_j)+m\left(\left<\theta_0,\mathcal{X}_i\right>\right)+\varepsilon_i, \label{mod_sim2}
	\end{equation}
	where, in this case:
	\begin{itemize}
		\item Ten non-null coefficients were considered, which correspond to the following impact points:
		
		\begin{minipage}[b]{.5\linewidth}
			\begin{eqnarray}
				\beta_{0j_1}&=&1.0,\quad    t_{j_1}=0.15, \nonumber\\
				\beta_{0j_2}&=&1.2, \quad  t_{j_2}=0.16,\nonumber\\
				\beta_{0j_3}&=&1.0,\quad    t_{j_3}=0.17, \nonumber\\
				\beta_{0j_4}&=&1.2, \quad  t_{j_4}=0.18, \nonumber\\
				\beta_{0j_5}&=&1.0,\quad    t_{j_5}=0.19, \nonumber
			\end{eqnarray} 	
		\end{minipage}\begin{minipage}[b]{.5\linewidth}
			\begin{eqnarray}
				\beta_{0j_6}&=&1.0, \quad t_{j_6}=0.70, \nonumber\\
				\beta_{0j_7}&=&1.2, \quad  t_{j_7}=0.71, \nonumber\\
				\beta_{0j_8}&=&-1.2,\quad t_{j_8}=0.72, \nonumber\\
				\beta_{0j_9}&=&-1.2,\quad t_{j_9}=0.73, \nonumber\\
				\beta_{0j_{10}}&=&-1.2, \quad t_{j_{10}}=0.74.\label{gip_m}
			\end{eqnarray} 
		\end{minipage}
		\item The curves involved in the nonlinear part were generated from (\ref{X-sim}), but the random variables $a_i$, $b_i$ and $c_i$  ($i=1,\dots,n$) are independent and uniformly distributed in the interval $[0,5]$.
	\end{itemize}
	Owing to (\ref{gip_m}), we obtained a Grouped-Impact-Point MFPLSIM, GIP-MFPLSIM (the relevant variables are consecutive in case $p_n=101$). A total of
	$M=100$ independent samples were generated from the GIP-MFPLSIM (\ref{mod_sim2}), which were divided into $\mathcal{D}_{train}$ (see (\ref{d_train})) and $\mathcal{D}_{test}$ (see (\ref{d_test})). Subsequently, FASSMR and IASSMR  were applied following the same scheme and considerations as the first scenario; however, here  we considered the same sample size for the two procedures (which is closer to real data applications). Then, in the case of IASSMR, the sample size was divided into two parts to be used in the first and second steps. In this application, we considered $n_1=n_2=n/2$.
	
	We compared the computational time required by both methods to estimate a sample, MSEP (\ref{MSEP-sim}) and precision of impact point selection.
	With respect to this last point, we considered the following sets:
	\begin{eqnarray}
		I_{n}=[0.14,0.20]\cup[0.69,0.75],\nonumber\\
		\overline{I}_{n}=[0,0.14)\cup(0.20,0.69)\cup(0.75,1],\nonumber
	\end{eqnarray}
	and %we will classify as well-chosen all those  selected points belonging to $I_{n}$ and as wrongly-chosen those belonging to $\overline{I}_{n}$. That is, 
	quantified  Right=$\sharp\{I_{n}\cap \widehat{T}_{n}\}$ and Wrong=$\sharp\{\overline{I}_{n}\cap \widehat{T}_{n}\}$ for IASSMR and FASSMR.

	\subsubsection{Results}\label{sim1-res2}

	Table \ref{CT_IASSMR_FASSMR} summarises  the computational time required by  IASSMR and FASSMR for estimating one sample; it is evident that the computational time required by IASSMR is affect by $p_n$ (also derived from Table \ref{TIME_FASSMR_IASSMR}), while that required by FASSMR is only affected by increasing $n$. It is noteworthy remark that  for a moderate sample size ($n=200,300$) and small $p_n$, computational time required by IASSMR is similar or even smaller than that required by FASSMR. This is due to the division of the sample in the IASSMR two-stage procedure.
	\begin{table}[htp]
		\centering
		\caption{\label{CT_IASSMR_FASSMR} Computational time in seconds needed for the estimation of one sample from the described modification of (\ref{mod_sim}), using IASSMR and FASSMR for different values of $n$ and $p_n$. The  values eligible for $w_n$ are those belonging to $W^*$ (see (\ref{W^*})). The results were obtained with a computer with the following features: Intel Core i7-7700HQ CPU, 8 GB RAM, 1 TB HDD, 256 GB SSD.}
		\scalebox{0.9}{
			\begin{tabular}{|lr|rrrrr|}
				\hline
				$n$&Method & $p_n=101$ & $p_n=201$ & $p_n=501$ & $p_n=1001$ & $p_n=10001$\\
				\hline
				\multirow{2}{1cm}{$100$}& FASSMR & 405.53& 479.6 & 436.22 
				& 260.44& 251.28\\
				&IASSMR& 653.26 &1156.32 &3016.60  & 4877.91 &24860.64  \\
				\hline
				\multirow{2}{1cm}{$200$}&	FASSMR & 983.36& 822.14& 805.41&580.97&558.92 \\
				&	IASSMR& 931.11 & 1070.92 &  3047.36& 5450.66& 31641.58\\
				\hline	
				\multirow{2}{1cm}{$300$}&	FASSMR&2241.66 &2080.84 &1979.23 &  2062.11& 2337.90\\
				&	IASSMR & 1684.22 &1950.67 & 2290.78 & 9041.99& 71789.74\\	
				\hline
		\end{tabular}}
	\end{table}

	Tables \ref{MSEP_IASSMR_FASSMR} and \ref{IP_FASSMR_IASSMR} allow us to analyse and compare the accuracy of predictions and variable selection, respectively, obtained by FASSMR and IASSMR.  Some general observations can be derived from those tables:
	\begin{enumerate}[i)]
		\item The performance of both procedures benefited from the increase in sample size ($n$). However, it is difficult to analyse the effect of increasing the number of linear covariates ($p_n$). Both procedures were adversely affected with the increase in $p_n$ if we compare $p_n=101$ with $p_n=201$ and $p_n=501$. However, from $p_n=201$ to $p_n=501$ there was no deterioration in results. Results for $p_n=501$ were even better in some cases. 
		\item  For a small sample size ($n=100$), FASSMR surpassed the results obtained by IASSMR
		in terms of MSEP for every considered value of $p_n$. In contrast, the number of variables
		correctly selected was larger in the case of IASSMR, along with the number of wrongly
		selected variables.
		\item  For a moderate sample size ($n=200$ and $n=300$), IASSMR yielded better results
		than FASSMR for all considered values of $p_n$. %\textcolor{blue}{NO ENTIENDO LO QUE QUIERES DECIR CON LA FARSAE SIGUIENTE}\textcolor{red}{This improvement is even more evident in the case of variables well and wrong selected}.
	\end{enumerate}
	
	\begin{table}[htp]
		\centering
		\caption{	\label{MSEP_IASSMR_FASSMR}For $M=100$ samples from the GIP-MFPLSIM (\ref{mod_sim2}), mean of MSEP obtained for FASSMR  and IASSMR procedures.}
		\scalebox{0.9}{
			\begin{tabular}{lr|rr|rr|rr|}
				\cline{3-8}
				&& \multicolumn{6}{|c|}{MSEP} \\
				\cline{3-8} 
				& &\multicolumn{2}{|c|}{$p_n=101$}&\multicolumn{2}{c|}{$p_n=201$}&\multicolumn{2}{c|}{$p_n=501$}\\
				\hline
				\multicolumn{1}{|l}{$n$}& Method & Mean & SD & Mean & SD& Mean & SD\\
				
				\hline	
				\multicolumn{1}{|c}{	\multirow{2}{1cm}{$100$}}&	FASSMR  &0.5827&0.3208 &  0.6908& 0.3890&0.6803&0.3743\\
				\multicolumn{1}{|c}{}	&	IASSMR  &  1.0988 &2.0134 &3.3929 &5.6060& 2.7154&4.9089 \\
				\hline
				\multicolumn{1}{|c}{	\multirow{2}{1cm}{$200$}}&	FASSMR & 0.4076&0.1484& 0.4579&0.1443 &0.4510&0.1625\\
				\multicolumn{1}{|c}{}	&IASSMR  &0.3954&0.2038  &0.4097 &0.2154& 0.4255&0.2522\\
				\hline
				\multicolumn{1}{|c}{	\multirow{2}{1cm}{$300$}}&	FASSMR &0.3573&0.1217&0.4127&0.1296 &0.3857&0.1074\\
				\multicolumn{1}{|c}{}	&IASSMR  & 0.2916&0.1208 &0.3142&0.1326& 0.3018&0.1166
				\\
				\hline
		\end{tabular}}
	\end{table}

	\begin{table}[htp]
		\centering
		\caption{	\label{IP_FASSMR_IASSMR} For $M=100$ samples from the GIP-MFPLSIM (\ref{mod_sim2}), the mean number of variables correctly selected ($\sharp\{I_{n}\cap \widehat{T}_{n}\}$) and incorrectly selected ($\sharp\{\overline{I}_{n}\cap \widehat{T}_{n}\}$) by FASSMR and IASSMR.}
		\scalebox{0.9}{
			\begin{tabular}{cc|cc|cc|cc|}
				\cline{3-8}
				&&\multicolumn{2}{c|}{$p_n=101$}&\multicolumn{2}{c|}{$p_n=201$}&\multicolumn{2}{c|}{$p_n=501$}\\
				\hline
				\multicolumn{1}{|l}{$n$}& Method&Right&Wrong&Right& Wrong&Right&Wrong\\
				\hline	
				\multicolumn{1}{|c}{	\multirow{2}{1cm}{$100$}}&	FASSMR &1.95&2.45 &1.78
				&2.81&1.76 &3.12\\
				\multicolumn{1}{|c}{}&	IASSMR  &4.11   &3.52 &4.26&8.34
				&4.69&8.85
				\\
				\hline
				\multicolumn{1}{|c}{	\multirow{2}{1cm}{$200$}}&	FASSMR &1.99& 2.06 & 1.90 &2.61&1.88&2.54
				\\
				\multicolumn{1}{|c}{}	&IASSMR  & 4.46 &1.12 &4.55&1.59&4.68&2.08 \\
				\hline
				\multicolumn{1}{|c}{\multirow{2}{1cm}{$300$}}&	FASSMR &2.00&2.08 &1.92&2.23&1.87
				&2.21
				\\
				\multicolumn{1}{|c}{}	&IASSMR  &4.82  & 0.50& 4.84&0.79&5.17&0.98 \\
				\hline
		\end{tabular}}
		
	\end{table}
It is worth noting that ii) is a consequence of dividing the sample of size $100$ into two subsamples of size $50$ to perform the two-stage procedure associated with IASSMR. This
sample size seems to be insufficient for obtaining a good estimation of  $\theta_0$.

	When the sample size was sufficient (observation iii), the second stage in IASSMR enabled us to recover some information that was lost in the first stage; in this case, the results
	by IASSMR were less affected by the discretization size and  $w_n$, hereby surpassing those
	obtained by FASSMR both in terms of MSEP and the number of correctly and incorrectly
	selected variables.

	\subsubsection{Final remarks and conclusions}
	The simulation study performed in the aforementioned two scenarios (spaced and
	grouped impact points) illustrates the utility of refining FASSMR to obtain a more sophisticated
	algorithm. In particular, as expected and was highlighted later by theoretical
	asymptotics (see result (\ref{gip})), the IASSMR procedure surpassed the drawbacks of FASSMR
	in terms of the selected impact points. This improvement accompanies high predictive performance
	and reasonable computational costs. Furthermore, the comparisons between the
	practical behavior of FASSMR and IASSMR in terms of the MSEP, accuracy of variable
	selection, and computational time provide practical guidelines about algorithms that can be
	used in each practical situation. Thus, the main recommendations can be summarised as
	follows:
	\begin{itemize}
		\item For a small $n$ and big $p_n$, we should use FASSMR.
		\item For a large/moderate $n$ and small/moderate $p_n$, IASSMR should be used. 
		\item For large $n$ and $p_n$, FASSMR provides an initial approximation; IASSMR yields precise
		set of selected variable but at a higher computational cost. However, the computational
		time required by IASSMR is significantly lower than that required by PLS methods.
	\end{itemize} 
	
	\section{Real data application}
	In this section, we discuss the usefulness of the proposed methodology through its
	application for solving a real problem: the prediction of ash content in a sample of sugar,
	with its absorbance spectra at two different excitation wavelengths.
	Although the ash content can be determined by chemical analysis, the use of functional
	regression for predicting it will be economically advantageous.
	Therefore, this section is devoted to the analysis of this dataset using the flexible model
	and algorithms presented in this paper.
	
	\subsection{Data}
	The data presented in Section 1.2 were analysed. Accordingly, we have $268$ samples from $(Y,\zeta,\mathcal{X})$, where $Y$ is a scalar random variable (ash content) and $\zeta$ and $\mathcal{X}$ are functional random variables (absorbance spectra from $275$ to $560nm$ at excitation wavelengths $240nm$ and $290nm$, respectively; both variables were observed on $p_n=571$ equally spaced wavelengths in the interval $[275,560]$).
	Although the number of available samples was $268$, two samples were discarded in this application as extreme outliers. Therefore, our dataset comprised $266$ samples $\mathcal{D}=\{(\zeta_i,\mathcal{X}_i,Y_i),\quad i=1,\dots,266\}$, and we aimed to predict $Y$ with $\zeta$ and $\mathcal{X}$. More details, including graphics of the curves, can be found in Section \ref{chemometrics}.
	
	To evaluate the models and estimation methods proposed in this study, 
	our dataset, $\mathcal{D}$, was split into two subsamples: the training sample $\mathcal{D}_{train}=\{(\zeta_i,\mathcal{X}_i,Y_i),\quad i=1,\dots,216\}$ and the testing sample $\mathcal{D}_{test}=\{(\zeta_i,\mathcal{X}_i,Y_i),\quad i=217,\dots,266\}$. Therefore, $\mathcal{D}_{train}$ was used for estimating, while $\mathcal{D}_{test}$ was used to measure the quality of predictions. Accordingly, we used the MSEP (\ref{MSEP-sim}) with $n=216$ and $n_{test}=50$ now.

	\subsection{Modelling step}
	To determine the initial effect of each functional variable in the response, we performed
	a preliminary study with modelling data using two unifunctional models: a functional linear
	model (FLM) and FSIM. In both cases, we constructed models with each functional variable.
	The FLM was estimated using principal component analysis (PCA) (via the {\tt fregre.pc} function in the \textsf{fda.usc R} package).
	The FSIM was estimated using Nadaraya-Watson type estimators (see \citealt*{ferpv03}) with Epanecknikov kernel, selecting $h$ using the BIC. $\theta$ was estimated using the procedure described in \cite*{novo_2019}, using $l=3$ and selecting $m_n$, see expression (\ref{theta_0_base}), using the cross-validation procedure (\citealt*{novo_2019}). 
	
	The models and results of MSEP are summarised in Table \ref{table_p_s2}. It can be observed that a linear effect is convenient for $\zeta$ (the lowest MSEP was obtained); however, to ensure that
	the same variable is not considered in both parts of the model, variable $\mathcal{X}$ should be entered semiparametrically.

	\begin{table}[htp]
		\centering
		\caption{	\label{table_p_s2}Unifunctional regression models and values of the criterion error.}
		\begin{tabular}{|ll|}
			\hline
			FLM &MSEP \\ 
			\hline 
			%&  \\ 
			$Y=a+\int_{275}^{560}\mathcal{X}(t)\alpha(t)dt+\varepsilon $& $4.588$\\
			$Y=a+\int_{275}^{560}\zeta(t)\alpha(t)dt+\varepsilon $& $2.207$\\
			%&\\
			\hline
			FSIM&MSEP\\
			\hline
			%&\\
			$Y=r\left( \left\langle \theta,\mathcal{X}\right\rangle \right)
			+\varepsilon $ & $3.6981$ \\
			$Y=r\left( \left\langle \theta,\zeta\right\rangle \right)
			+\varepsilon $ & $2.6802$\\%&\\
			\hline		
		\end{tabular}
		
	\end{table}
	
	Therefore, the following model can be considered:
	\begin{equation}
		Y_i=\sum_{j=1}^{571}\beta_{0j}\zeta_i(t_j)+m\left(\left<\theta_0,\mathcal{X}_i\right>\right)+\varepsilon_i, \label{mod_data}
	\end{equation}
	and the results obtained by combining this model with the proposed variable selection
	methods can be analysed.

	\subsection{Results}
	
	To estimate model (\ref{mod_data}), we applied the standard PLS method, FASSMR, and IASSMR.
	For this task, with respect to the four cases, the technical considerations derived in \ref{T_C} were used. Value $l=3$ was considered in (\ref{theta_0_base}), while $m_n$ was selected via the BIC criterion.  
	
	\begin{table}[htp]
		\centering
		\caption{\label{table_sugar} Results obtained applying each variable selection method to model (\ref{mod_data}).}%	
		\begin{tabular}{c|rrrrr|}
			\cline{2-6}
			&MSEP&$\widehat{s}_n$&	$\widehat{m}_n$& $\widehat{w}_n$& $p_t$ \\
			\hline
			
			\multicolumn{1}{|c|}{PLS} & 6.0375 &  53  & 4&-& 9.8689 \\
			\multicolumn{1}{|c|}{FASSMR}&3.0329&8&2&15& 1\\
			\multicolumn{1}{|c|}{IASSMR}& 2.0064& 9 &2&15 & 4.3399\\
			\hline
		\end{tabular}	
	\end{table}

	\begin{figure}[htp]
		\centering
		\makebox{\includegraphics[width=0.66\textwidth]{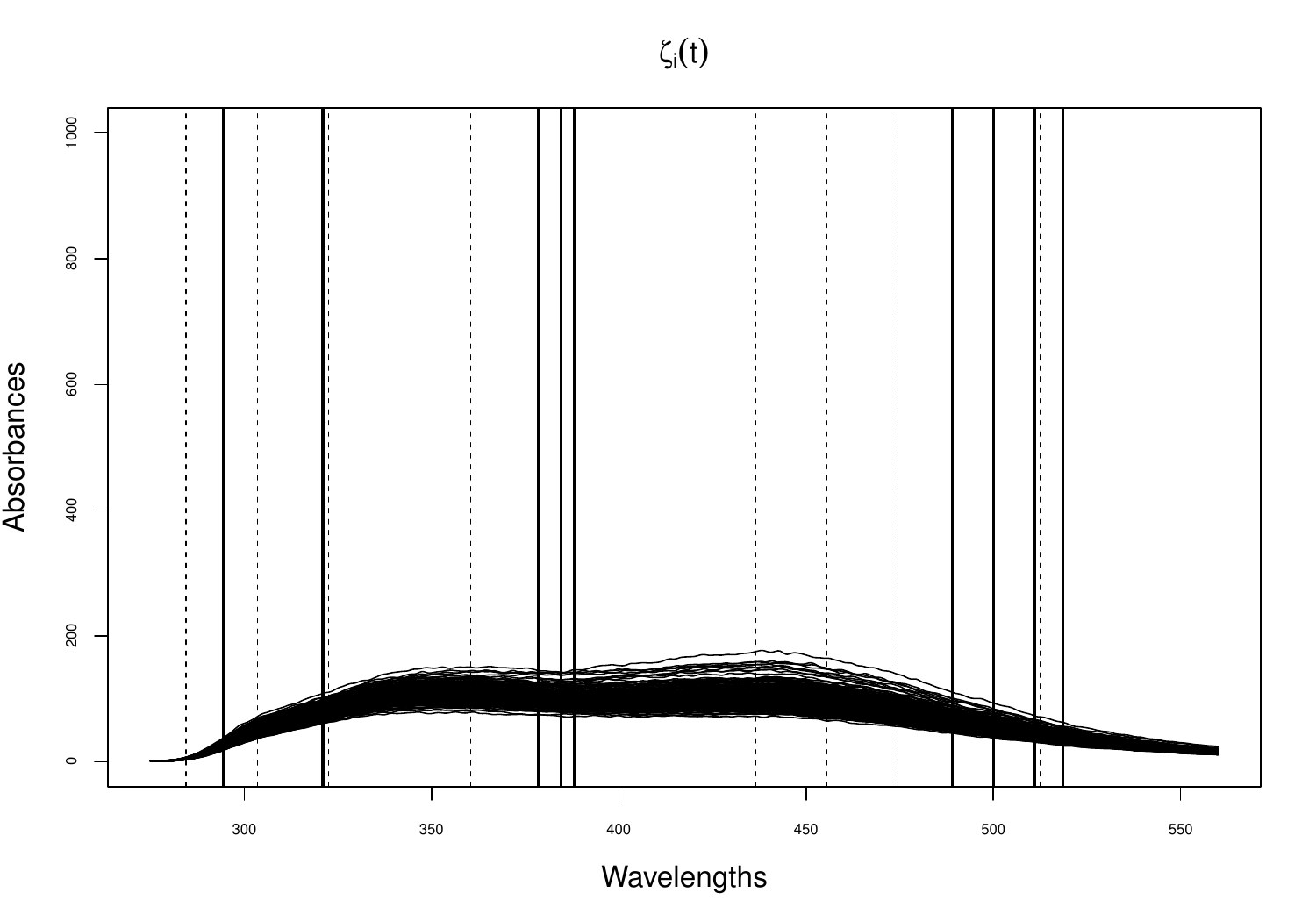}}
		\caption{\label{fig4}
			Absorbance curves at excitation wavelengths  $240nm$ ($\mathcal{\zeta}$) with selected impact points using the FASSMR (dashed vertical lines) and the IASSMR (solid vertical lines).}
	\end{figure}
	\ref{table_sugar} lists the numerical results; column $p_t$ contains the proportion of time required
	by the three methods for yielding the final results in comparison with the fastest algorithm
	among the three (FASSMR). The PLS method offered the most complex model with a total
	of $53$ linear covariates and a complicated expression for the estimated direction $\widehat{\theta}_0$ (four regularly spaced interior knots were required for its B-spline representation). Furthermore,
	this complexity was accompanied by the worst results in terms of MSEP and computational
	time. In contrast, FASSMR evidently improved the PLS results in terms of complexity of
	model (yielding simpler model) and MSEP; however, the best result in terms of MSEP was
	obtained by IASSMR. This is related to set $\widehat{S}_n$ obtained with this algorithm: the second
	stage in IASSMR specifies and completes the set of relevant variables provided by FASSMR. Figure \ref{fig4} illustrates this and demonstrates some GIP structure can be present around $385nm$; in addition, it illustrates that in this example, none of the relevant variables selected by
	FASSMR were selected by IASSMR (note that this is not incoherent because the objective
	of IASSMR is to refine the selection made by FASSMR). We should also note that IASSMR
	is faster than the PLS procedure.

\begin{figure}[htp]
		\centering
		\makebox{\includegraphics[width=0.33\textwidth]{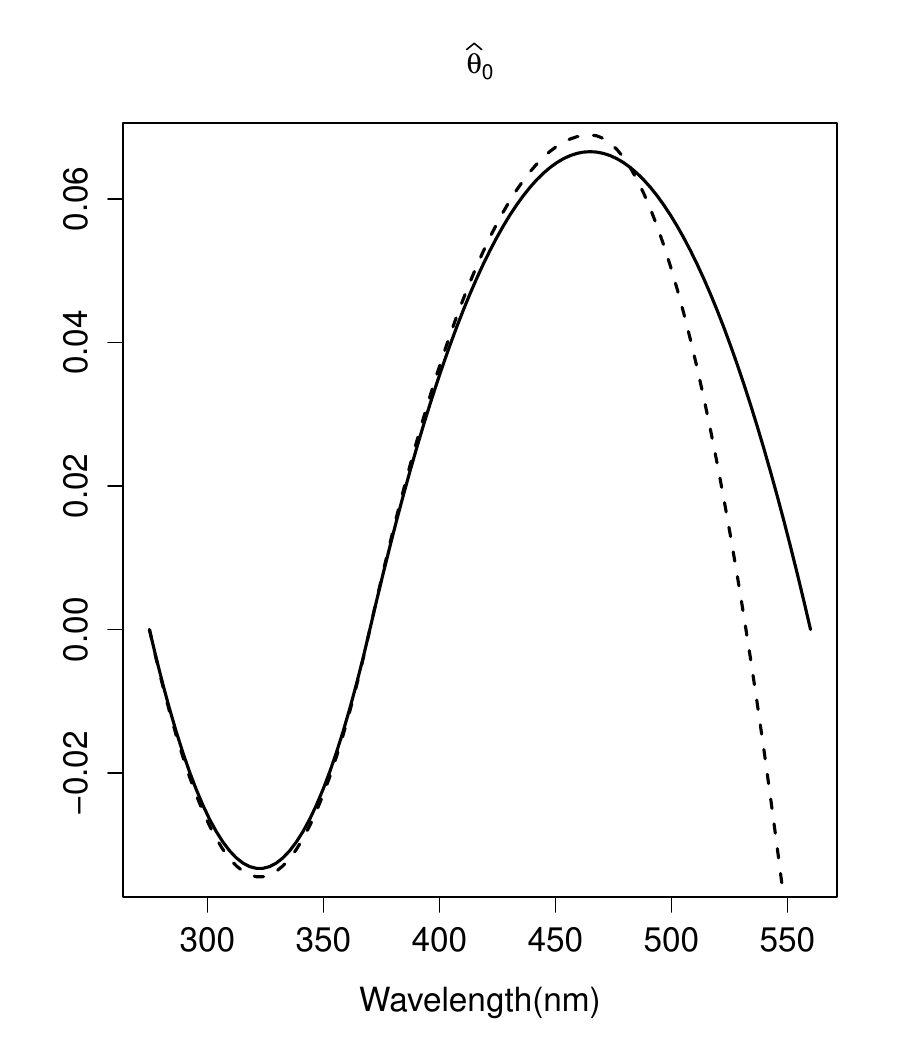}}
		\caption{\label{fig5}
			Estimated direction using IASSMR (solid line) and FASSMR (dashed line).}
	\end{figure}

	Finally, as evident from Figure \ref{fig5}, the estimated direction using both algorithms exhibited
	a similar shape; in both cases, it denotes a bump around $325nm$ and a peak around $475nm$, which could be important indicators of the effect of $\mathcal{X}$ on the ash content of sugar.

	\subsection{Summary of conclusions}
The sugar data application illustrates the usefulness of MFPLSIM in modelling real
problems and the good performance of the presented algorithms in estimating this model.
MFPLSIM has two significant advantages: it allows the inclusion of more than one functional
covariate in the model and these covariates enter with interpretable parameters ($\pmb{\beta}_0$ and $\theta_0$). In addition, its semiparametric feature, combined with a good estimation tool, provides low
prediction errors. Furthermore, the developed algorithms for variable selection and estimation of MFPLSIM demonstrated good performance in comparison with the standard PLS method
in terms of the MSEP and computational time. FASSMR provides quick preliminary results,
while IASSMR provides refined estimations. In particular, the combination of MFPLSIM
and IASSMR seems to be a potent tool because it achieved the best result in terms of the
MSEP.

	\appendix
	
	\section{Theoretical issues}
	\subsection{Asymptotics for the FASSMR algorithm}
	To present some theoretical results related to variable selection performed by FASSMR,
	we include some technical assumptions: 
	\begin{description}
		\item[Conditions on the curve $\zeta$.] The curve $\zeta$ is observed in a grid such that
		\begin{equation}
			\textrm{exist } c_1,c_2, \ \ \textrm {for all }j=1,\dots,p_n-1,\ \ \textrm{such that} \ \ 0<\frac{c_1}{p_n}<t_{j+1}-t_j<\frac{c_2}{p_n}<\infty.
			\label{dis}
		\end{equation}
		In addition,
		\begin{equation}
			\zeta \textrm{ is Lipschitz continuous on its support, }
			\label{lipschitz}
		\end{equation}
		and bounded away from zero; that is:
		\begin{equation}
			\textrm{exists  } c, \ \  \textrm{for all } t\in [a,b], \ \ \textrm{such that} \ \ \left|\zeta(t)\right|\geq c>0.
			\label{etiqueta-olvidada}
		\end{equation}
		\item[Conditions on the coefficients of the model.] Let us assume that
		\begin{eqnarray}
			\textrm{exists  } c, \ \  \textrm{for all }  j=1,\dots,q_n,\ \ \textrm{for all } k=1,\dots, w_n, \ \ \textrm{such that} \nonumber\\  \beta_{0j+(k-1)q_n}\not =0 \Longrightarrow \left|\sum_{j=1}^{q_n}\beta_{0 j+(k-1)q_n}\right|>c>0.
			\label{s1}
		\end{eqnarray}
		\item[Conditions on the standard variable selection method.]
		Specifically, let us consider SFPLSIM:
		\begin{equation}
			Y=\sum_{k=1}^{w_n}\alpha_{0j}X_j+g\left(\left<\delta_{0},\mathcal{X}\right>\right)+\varepsilon,
			\label{sfplsim1}
		\end{equation}
		where $X_j$ denotes random real covariates, $\delta_0$ is an unknown functional direction; moreover,
		if $S_{n}^*=\left\{k=1,\dots,w_n,\ \alpha_{0j}\not=0\right\}$ and $\sharp(S_n^*)=s_n^*$, it is verified that
		$s_n^*=o(w_n)$.
		The standard SCAD-PLS procedure leads to estimates $\widehat{\alpha}_{0j}$ of $\alpha_{0j}$ satisfying the following property:
		\begin{eqnarray}
			\Pr\left(\{k=1,\dots,w_n, \textrm{ such that }\alpha_{0k}=0\}=\{k=1,\dots,w_n, \textrm{ such that } \widehat{\alpha}_{0k}=0\}\right)\rightarrow 1,\nonumber\\ \textrm{ when } n\rightarrow\infty.\label{oracle_scad1}
		\end{eqnarray}
	\end{description}
	\begin{remark}
		Note that the suitable conditions under which (\ref{oracle_scad1}) holds true can be found in \cite*{novo}. Moreover, assumption(\ref{s1}) is specific to the functional setting
		addressed in this paper (scalar variables with functional origin). This condition (\ref{s1})  was first introduced in \cite{aneirosv_2014}; discussion and examples under which this condition is satisfied can be found in \cite{aneirosv_2014,aneirosv_2015}.
	\end{remark}

	Finally, to introduce the  theoretical result,  for each $j=1,\dots, p_n$,  $k_j$ denotes the unique integer $k\in\{1,\dots,w_n\}$ such that $j\in\{(k-1)q_n+1,\dots,kq_n\}$. The following result establishes the relationship between the variable selection procedures related to models (\ref{model})  and (\ref{mod_red}) as follows: if the $j$th ($j=1,\dots,p_n$) variable is relevant in model (\ref{model}), for a sufficient $n$, the corresponding $k_j$th neighbouring variable  in model (\ref{mod_red}) will be estimated to be non-null; conversely, if the $k$th variable is estimated as non-null in model (\ref{mod_red}), there will be a relevant neighbouring variable in model (\ref{model}).
	\begin{proposition}
		Under conditions (\ref{model})--(\ref{sample}), (\ref{mod_red}), (\ref{iv3}), (\ref{dis})--(\ref{oracle_scad1}), assuming that $w_n\rightarrow\infty$ when $n\rightarrow\infty$, it is verified that:

		\begin{enumerate}
			\item $\Pr\left(\textrm{for all } j\in S_n,\ \widehat{\beta}_{0k_j}^{\pmb{1}}\not=0\right)\longrightarrow 1$ when $n\rightarrow\infty$.
			\item  $\Pr\left(\textrm{for all } k\in \widehat{S}_n^{\pmb{1}},\ \textrm{exists } j \in \{1,\dots,q_n\} \textrm{ such that }\beta_{0j+(k-1)q_n}\not=0\right)\longrightarrow 1$ when $n\rightarrow\infty$.
		\end{enumerate}
		\label{proposition}
	\end{proposition}
	
	\subsection{Asymptotics for IASSMR}
To introduce some asymptotic results related to the estimators derived from IASSMR,
let us add some hypotheses to those required for FASSMR:
	\begin{description}
		\item[Conditions on the coefficients of the model.]
		\begin{eqnarray}
			\textrm{For all } k=1,\dots, w_n, \ \ \textrm{exists } 0<a_k<\infty, \ \ \textrm{such that } \sum_{j=1}^{q_n}\beta_{0 j+(k-1)q_n}\not =0\Longrightarrow \nonumber \\\sharp S^k \sim a_kq_n \ \ as \ \  n\rightarrow \infty,\label{s2}
		\end{eqnarray}
		where for any $k=1,\dots,w_n,$
		\begin{equation*}
			S^k=\left\{j=1,\dots,p_n, \textrm{ such that }  j=1+ (k-1)q_n,\dots,kq_n \textrm{ and } \beta_{0j}\not =0\right\}.
		\end{equation*}
		\item[Conditions on the standard variable selection method.]
		Let us consider SFPLSIM
		\begin{equation}
			Y=\sum_{j\in\mathcal{P}_n}\alpha_{0j}X_j+g\left(\left<\delta_{0},\mathcal{X}\right>\right)+\varepsilon,\label{sfplsim2}
		\end{equation}
		where $\mathcal{P}_n\subset\{1,\dots,p_n\}$ with $\sharp\mathcal{P}_n=O(w_n)$ or $\sharp\mathcal{P}_n=O(s_n)$. The standard SCAD-PLS procedure leads to estimates $\widehat{\alpha}_{0j}$ of $\alpha_{0j}$ and $\widehat{\delta}_0$ of $\delta_0$ satisfying properties:
		\begin{equation}
			\Pr\left(\{j\in\mathcal{P}_n,\ \alpha_{0j}=0\}=\{j\in\mathcal{P}_n,\ \widehat{\alpha}_{0j}=0\}\right)\longrightarrow 1,\textrm{ as } n\rightarrow\infty,\label{oracle_scad}
		\end{equation}
		\begin{equation}
			\textrm{there exists }\gamma\geq 0 \textrm{ such that }\left\lvert\left\lvert\widehat{\pmb{\alpha}}_0-\pmb{\alpha}_0\right\lvert\right\lvert=O_p\left(n^{-1/2}\left(\sharp\mathcal{P}_n \right)^{\gamma}\right)\label{rate_beta_scad}
		\end{equation}
		\begin{equation}
			\textrm{ and } \textrm{there exists } d:\mathbb{R}\rightarrow (0,\infty) \textrm{ such that }	\left\lvert\left\lvert\widehat{\delta}_0-\delta_0\right\lvert\right\lvert=O_p\left(n^{-1}d(h)\left(\sharp\mathcal{P}_n \right)^{\gamma-3/2} \right),\label{rate_theta_scad}
		\end{equation}
		where $\pmb{\alpha}_0=(\alpha_{0j}, \ j\in\mathcal{P}_n )^{\top}$,  $\widehat{\pmb{\alpha}}_0=(\widehat{\alpha}_{0j}, \ j\in\mathcal{P}_n )^{\top}$. 
		\item[Conditions on the semiparametric estimate.]  Let us consider the following semiparametric models:
		\begin{equation*}
			Y=g_0\left(\left<\delta_0,\mathcal{X}\right>\right)+\varepsilon,
		\end{equation*}
		\begin{equation*}
			X_j=g_j\left(\left<\delta_0,\mathcal{X}\right>\right)+\eta_j, \quad j=1,\dots, p_n,
		\end{equation*}
		and denote
		$g_{j,\delta_0}(\chi)\equiv g_j\left(\left<\delta_0,\chi\right>\right)$ con $j=0,\dots,p_n$.
		Let $\widehat{g}_{j,\delta_0}(\chi)$ be the corresponding semiparametric estimate for $g_{j,\delta_0}(\chi)$ $j=0,\dots,p_n,$ from the aforementioned models
		using the same types of weights used in IASSMR, and $\delta\in\Theta_n\subset\mathcal{H}$. The following assumptions are required:
		\begin{equation}
			\sup_{\delta \in \Theta_n} \sup_{\chi\in \mathcal{C}}\left\{ \left|\widehat{g}_{0,\delta}(\chi)-g_{0, \delta_0}(\chi)\right|\right\} =O_{p}\left(a_n \right),\label{cond_m1}
		\end{equation}
		\begin{equation}
			\max_{j\in S_n} \sup_{\delta \in \Theta_n} \sup_{\chi\in \mathcal{C}}\left\{ \left|\widehat{g}_{j,\delta}(\chi)-g_{j, \delta_0}(\chi)\right|\right\} =O_{p}\left(b_n \right),\label{cond_m2}
		\end{equation}
		\begin{equation}
			\max_{j\in S_n}\sup_{\chi\in \mathcal{C}}\left\{ \left|g_{j, \delta_0}(\chi)\right|\right\} =O\left(1 \right).\label{cond_m3}
		\end{equation}
	\end{description}
	\begin{remark} Condition (\ref{s2}) is specific to the framework of scalar variables with functional
		origin (for details, discussion and examples under which this condition is satisfied, see \citealt*{aneirosv_2014,aneirosv_2015}).
		In addition, \cite*{novo} stated the conditions under which (\ref{oracle_scad})--(\ref{cond_m3}) hold true, including the characterization of the function $d$ and the functional subset $\Theta_n$.
		Similarly, Lemma 6 in \cite*{novo} specifies rates $a_n$ and $b_n$. 
	\end{remark}

The next theorem presents some asymptotic results related to our proposed estimators
obtained from IASSMR. For the expressions of  $\widehat{S}_n$, $\widehat{\pmb{\beta}}_0$ and $\widehat{\theta}_0$, see (\ref{S_n_est}), (\ref{est_beta0}) and (\ref{est_theta0}), respectively.
	
	\begin{theorem}\label{th2}
		Under conditions (\ref{model})--(\ref{sample}), (\ref{iv3}), (\ref{dis})--(\ref{s1}), (\ref{s2})--(\ref{rate_theta_scad}), and if $w_n\rightarrow\infty$ as $n\rightarrow\infty$, the following is obtained
		\begin{equation}
			\left|\left|\widehat{\pmb{\beta}}_0-\pmb{\beta}_0\right|\right|=O_p\left(n^{-1/2}s_{n}^{\gamma}\right),\label{res_beta}
		\end{equation}
		\begin{equation}
			\left|\left|\widehat{\theta}_0-\theta_0\right|\right|=O_p\left(n^{-1}d(h) s_{n}^{\gamma-3/2}\right),\label{res_theta}
		\end{equation}
		and 
		\begin{equation}
			\Pr\left(\widehat{S}_n=S_n\right)\longrightarrow 1,\quad n\rightarrow\infty.\label{res_S_n}
		\end{equation}
	\end{theorem}
	
	Finally, using the  estimation of linear coefficients obtained in (\ref{est_beta0}), for each $\theta\in\mathcal{H}$, we define
	\begin{equation*}
		\widehat{m}_{\theta}(\chi)\equiv \widehat{m}\left(\left<\theta,\chi\right>\right)=
		\frac{\sum_{i=1}^n\left(Y_i-\pmb{\zeta}_i^{\top}\widehat{\pmb{\beta}}_0\right) K\left(d_{\theta}\left(\chi,\mathcal{X}_i\right)/h\right) }{\sum_{i=1}^nK\left(d_{\theta}\left(\chi,\mathcal{X}_i\right)/h\right)}, \ \forall \chi \in \mathcal{H}.\label{m_theta}
	\end{equation*}
	
	\begin{theorem}
		Under assumptions of Theorem \ref{th2}, if conditions (\ref{cond_m1}),  (\ref{cond_m2}) and (\ref{cond_m3}) are satisfied,  $h\rightarrow 0$ and $b_n\rightarrow0$ as $n\rightarrow\infty$,  then
		\begin{equation}
			\sup_{\theta\in \Theta_n} \sup_{\chi\in \mathcal{C}}\left\{ \left|\widehat{m}_{\theta}(\chi)-m_{\theta_0}(\chi)\right|\right\} =O_{p}\left(a_n \right)+O_p\left(n^{-1/2}s_{n}^{\gamma+1/2}\right).\label{res_m}\end{equation}\\
		\label{th3}
	\end{theorem}
	\begin{corollary}
		Under assumptions of Theorem \ref{th3},  if $\Theta_n^{\pmb{2}}\subset\Theta_n$ and $\widehat{\theta}_0$ is the estimator of $\theta_0$  obtained in (\ref{est_theta0}), we have: 
		\begin{equation}
			\sup_{\chi\in \mathcal{C}}\left\{ \left|\widehat{m}_{\widehat{\theta}_0}(\chi)-m_{\theta_0}(\chi)\right|\right\} =O_{p}\left(a_n \right)+O_p\left(n^{-1/2}s_{n}^{\gamma+1/2}\right).\end{equation}
	\end{corollary}
	
	\subsubsection{The grouped impact point case}
	In FDA, the continuity of the curve $\zeta$ may let us expect that the impact points are grouped; that is, the relevant variables are very close on the discretisation. In the situation of Grouped-Impact-Point MFPLSIM (GIP-MFPLSIM), the introduction of the following conditions could seem valid:
	\begin{description}
		\item[Conditions on the grouping of the impact points.] There exist some intervals $I_1,\dots, I_{J_n}$ such that $I_i\cap I_j=\emptyset$ and such that $T_n\subset I_n$ (see definition of $T_n$ in (\ref{T_n})) where $I_n=\cup_{j=1}^{J_n}I_j$ and 
		\begin{equation}
			\Pr\left(T_n=I_n\right)\longrightarrow1 \textmd{ when } n\rightarrow\infty.\label{H}
		\end{equation}
	\end{description}
	
	\begin{corollary}
		Under the same conditions of Theorem \ref{th2}, if, in addition,  assumption (\ref{H}) holds, then
		\begin{equation}
			\Pr\left(\widehat{T}_n=I_n\right)\longrightarrow1 \textmd{ when } n\rightarrow\infty.\label{gip}
		\end{equation}
	\end{corollary}

	\subsection{Sketch form of the proofs}
	\emph{\large Proof of Proposition \ref{proposition}:}\\

	Note that assertion 1. of Proposition \ref{proposition} can be proved ensuring that
	\begin{eqnarray}
		\Pr\left(\textrm{exists } j\in S_n \textrm{ such that }  \widehat{\beta}_{0k_j}^{\pmb{1}}=0\right)&=&\nonumber\\
		\Pr\left(\textrm{exists } j=1,\dots,p_n, \ \beta_{0j}\not=0 \textrm{ such that }  \widehat{\beta}_{0k_j}^{\pmb{1}}=0\right)&\longrightarrow& 0 \textrm{ when } n\rightarrow \infty.\label{pprop1}
	\end{eqnarray}
	It should be noted that we are under the assumptions of Lemma 2 presented in \cite{aneirosv_2014}. Using the first assertion of that lemma, we obtain:
	\begin{eqnarray}
		\Pr\left(\textrm{exists } j=1,\dots,p_n, \ \beta_{0j}\not=0 \textrm{ such that }  \widehat{\beta}_{0k_j}^{\pmb{1}}=0\right)\leq\nonumber\\ \Pr\left(\textrm{exists } k=1,\dots,w_n, \ \beta_{0k}\not=0 \textrm{ such that }  \widehat{\beta}_{0k}^{\pmb{1}}=0\right).\label{pprop2}
	\end{eqnarray}
	Now using assumption (\ref{oracle_scad1}), the right hand term of expression (\ref{pprop2}) tends to $0$ as $n$ tends to $\infty$. Then, (\ref{pprop1})  is proved; consequently, assertion 1. of Proposition \ref{proposition} is also proved.
	
	Following analogous reasoning, assertion 2. of Proposition \ref{proposition} can be proved ensuring that 
	\begin{eqnarray}
		\Pr\left(\textrm{exists } k\in \widehat{S}_n^{\pmb{1}} \textrm{ such that for all }  j\in\{1,\dots,q_n\}, \ \beta_{0j+(k-1)q_n}=0\right)&=&\nonumber\\
		\Pr\left(\textrm{exists }  k=1,\dots,w_n, \ \widehat{\beta}_{0k}^{\pmb{1}}\not=0 \textrm{ such that for all }  j\in\{1,\dots,q_n\},  \ \beta_{0j+(k-1)q_n}=0\right)&\longrightarrow& 0 \nonumber\\ \textrm{ when } n\rightarrow \infty.\nonumber\\\label{pprop3}
	\end{eqnarray}
	Using the assumption (\ref{oracle_scad1}), we obtain:
	\begin{eqnarray}
		\Pr\left(\textrm{exists }  k=1,\dots,w_n, \ \widehat{\beta}_{0k}^{\pmb{1}}\not=0 \textrm{ such that for all }  j\in\{1,\dots,q_n\},  \ \beta_{0j+(k-1)q_n}=0\right)\leq \nonumber\\ \Pr\left(\textrm{exists }  k=1,\dots,w_n, \ \beta_{0k}^{\pmb{1}}\not=0 \textrm{ such that for all }  j\in\{1,\dots,q_n\},  \ \beta_{0j+(k-1)q_n}=0\right)+o(1).\nonumber\\\label{pprop4}
	\end{eqnarray}
	Now applying the second assertion in Lemma 2 presented in \cite{aneirosv_2014}, the right hand term in (\ref{pprop4}) tends to $0$ as $n$ tends to $\infty$. Therefore, (\ref{pprop3}) is proved; consequently assertion 2. of Proposition \ref{proposition} is proved.\\
	
	\emph{\large Proof of (\ref{res_beta}):}\\
	
	Defining 
	\begin{eqnarray}
		\mathcal{R}_n^{\pmb{1}*}=\left\{j=1,\dots,p_n, \textrm{ such that } \zeta(t_j)\in \mathcal{R}_n^{\pmb{1}}\right\}, \nonumber\\
		\mathcal{R}_n^{\pmb{2}*}=\left\{j=1,\dots,p_n, \textrm{ such that } \zeta(t_j)\in \mathcal{R}_n^{\pmb{2}}\right\},\nonumber
	\end{eqnarray}
	one can write:
	
	\begin{equation}
		\left|\left|\widehat{\pmb{\beta}}_0-\pmb{\beta}_0\right|\right|^2=\sum_{j\in \mathcal{R}_n^{\pmb{2}*}}\left(\widehat{\beta}_{0j}^{\pmb{2}}-\beta_{0j}^{\pmb{2}}\right)^2+\sum_{j\not\in \mathcal{R}_n^{\pmb{2}*}, j\in S_n}(\widehat{\beta}_{0j}-\beta_{0j})^2.\label{pbeta1}
	\end{equation}
	If we consider $\mathcal{P}_n=\mathcal{R}_n^{\pmb{2}*}$ in (\ref{rate_beta_scad}) and consider $\sharp\mathcal{R}_n^{\pmb{2}*}=O(s_n)$, using the reasoning employed to achieve (A.8) in \cite{aneirosv_2014} we obtain
	\begin{equation}
		\sum_{j\in \mathcal{R}_n^{\pmb{2}*}}\left(\widehat{\beta}_{0j}^{\pmb{2}}-\beta_{0j}^{\pmb{2}}\right)^2=O_p\left(n^{-1}s_n^{2\gamma}\right).\label{pbeta2}
	\end{equation}
	In addition, if we consider $\mathcal{P}_n=\mathcal{R}_n^{\pmb{1}*}$ in (\ref{rate_beta_scad}) and  use assumption (\ref{oracle_scad}) and the first assertion in Proposition \ref{proposition}, we obtain
	\begin{equation}
		\sum_{j\not\in \mathcal{R}_n^{\pmb{2}*}, j\in S_n}\left(\widehat{\beta}_{0j}-\beta_{0j}\right)^2=O_p\left(n^{-1}s_n^{2\gamma}\right).\label{pbeta3}
	\end{equation}
	(For specific details, see proof of (A.12) in \citealt*{aneirosv_2014}).
	Therefore, result (\ref{res_beta}) is obtained by combining (\ref{pbeta1}) with (\ref{pbeta2}) and (\ref{pbeta3}).\\

	\emph{\large Proof of (\ref{res_theta}):}\\
	
	\begin{equation}
		\left\lvert\left\lvert\widehat{\theta}_0-\theta_0\right\lvert\right\lvert=\left\lvert\left\lvert\widehat{\theta}_0^{\pmb{2}}-\theta_0\right\lvert\right\lvert\leq \left\lvert\left\lvert\widehat{\theta}_0^{\pmb{2}}-\theta_0^{\pmb{2}}\right\lvert\right\lvert+\left\lvert\left\lvert\theta_0^{\pmb{2}}-\theta_0\right\lvert\right\lvert\label{ptheta1}
	\end{equation}
	
If we consider $\mathcal{P}_n=\mathcal{R}_n^{\pmb{2}*}$ in (\ref{rate_theta_scad}) and use $\sharp\mathcal{R}_n^{\pmb{2}*}=O(s_n)$ (see (A.9) and (A.10) in \cite{aneirosv_2014}), we obtain
	\begin{equation}
		\left\lvert\left\lvert\widehat{\theta}_0^{\pmb{2}}-\theta_0^{\pmb{2}}\right\lvert\right\lvert=O_p\left(n^{-1}d(h) s_{n}^{\gamma-3/2}\right)\label{ptheta2}.
	\end{equation}
	
	Moreover, note that $\theta_0^{\pmb{2}}$ depends on the variable selection in the first stage. Then, we can ensure for all $\eta>0$ that
	\begin{eqnarray}
		\Pr\left(\left\lvert\left\lvert\theta_0^{\pmb{2}}-\theta_0\right\lvert\right\lvert\geq\eta n^{-1}d(h) s_{n}^{\gamma-3/2}\right)&\leq& \Pr\left(\left\lvert\left\lvert\theta_0^{\pmb{2}}-\theta_0\right\lvert\right\lvert\geq 0\right)\nonumber\\&=&\Pr\left(\textrm{exists } j\in S_n \textrm{ and } j\not\in\mathcal{R}_n^{\pmb{2}*}\right)\nonumber\\
		&\leq&\Pr\left(\textrm{exists } j=1,\dots,p_n, \ \beta_{0j}\not=0 \textrm{ and } \widehat{\beta}_{0k_j}^{\pmb{1}}=0\right).\nonumber
	\end{eqnarray}
	Therefore, if we use the first assertion in
	Proposition \ref{proposition}, we obtain for all $\eta>0$
	\begin{equation}
		\Pr\left(\left\lvert\left\lvert\theta_0^{\pmb{2}}-\theta_0\right\lvert\right\lvert \geq\eta n^{-1}d(h) s_{n}^{\gamma-3/2}\right)\rightarrow 0 \textrm{ as } n\rightarrow \infty.\label{ptheta3}\end{equation}
	Then, the desired result is obtained from the combination of (\ref{ptheta2}) and (\ref{ptheta3}) in (\ref{ptheta1}).\\

	\emph{\large Proof of (\ref{res_S_n}):}\\
	
	Similar to \cite{aneirosv_2014}, we can make the following decomposition:
	\begin{eqnarray}
		\Pr\left(\widehat{S}_n\not=S_n\right)&\leq&\Pr\left(\textrm{exists } j\in \mathcal{R}_n^{\pmb{2}*},\ \beta_{0j}^{\pmb{2}}\not=0\textrm{ and }  \widehat{\beta}_{0j}^{\pmb{2}}=0\right)+\Pr\left(\textrm{exists } j\in \widehat{\mathcal{S}}_{n}^{\pmb{2}}, \ \beta_{0j}^{\pmb{2}}=0\right) \nonumber\\
		&+&\Pr\left(\textrm{exists } j=1,\dots,p_n, \ \beta_{0j}\not=0\textrm{ and }  \widehat{\beta}_{0k_j}^{\pmb{1}}=0\right),
		\label{pS1}
	\end{eqnarray}
	where $\widehat{\mathcal{S}}_{n}^{\pmb{2}}=\{j\in\mathcal{R}_n^{\pmb{2}*}, \ \widehat{\beta}_{0j}^{\pmb{2}}\not=0\}$.
	
	If we consider $\mathcal{P}_n=\mathcal{R}_n^{\pmb{2}*}$ in (\ref{oracle_scad}), the  first two terms in the right hand side of (\ref{pS1}) tend to zero as $n\rightarrow\infty$.
	
	In contrast,
	if we apply the first assertion in Proposition \ref{proposition}, the  third term in the right hand side of (\ref{pS1}) tends to zero as $n\rightarrow\infty$.
	
	Consequently, $\Pr\left(\widehat{S}_n\not=S_n\right)\rightarrow 0$ as $n\rightarrow\infty$, and we obtain (\ref{res_S_n}).\\

	\emph{\large Proof of (\ref{res_m}):}\\
	
	The following can be easily obtained:
	\begin{eqnarray}
		\left|\widehat{m}_{\theta}(\chi)-m_{\theta_0}(\chi)\right|\leq \left|\widehat{g}_{0\theta}(\chi)-g_{0\theta_0}(\chi)\right|+\left(\sharp\left(S_n \cup \widehat{S}_n\right)\right)^{1/2}\times \nonumber\\\left(\sup_{u\in \mathcal{C}, j\in S_n \cup \widehat{S}_n, \theta \in \Theta_n}\left|\widehat{g}_{j\theta}(u)-g_{j\theta_0}(u)\right|\left|\left|\widehat{\pmb{\beta}}_0-\pmb{\beta}_0\right|\right|
		+ \sup_{u\in \mathcal{C}, j\in S_n \cup \widehat{S}_n}\left|g_{j\theta_0}(u)\right|\left|\left|\widehat{\pmb{\beta}}_0-\pmb{\beta}_0\right|\right|\right)\label{pm1}
	\end{eqnarray}
	
If we use conditions (\ref{number_imp_points}) and (\ref{res_S_n}), we obtain
	\begin{equation}
		\sharp\left(S_n \cup \widehat{S}_n\right)=O_p(s_n).\label{pm2}
	\end{equation}
In contrast, using conditions (\ref{cond_m3}) and (\ref{res_S_n}), the following is obtained:
	\begin{equation}
		\sup_{u\in \mathcal{C}, j\in S_n \cup \widehat{S}_n}\left|g_{j\theta_0}(u)\right|=O_p(1).\label{pm3}
	\end{equation}
	
	Consequently, using expressions (\ref{pm1})-(\ref{pm3}) and result (\ref{res_beta}), with conditions (\ref{cond_m1}), (\ref{cond_m2}) and assumptions $b_n\rightarrow 0$ and $h\rightarrow 0$  as $n\rightarrow \infty$, the desired result  (\ref{res_m}) can be obtained.\\

	\emph{\large Proof of (\ref{gip}):}\\
	
	It is a direct consequence of (\ref{res_S_n}).


\newpage
	
	
	
	\begin{thebibliography}{1}
		
		\bibitem[Ait-Sa\"idi et al.(2008)Ait-Sa\"idi, Ferraty, Kassa \& Vieu]{ait} AIT- SA\"IDI, A., FERRATY, F., KASSA, R. \& VIEU, P. (2008). Cross-Validated Estimations in the Single-Functional Index Model,  \emph{Statistics} \pmb{42}(6), 475--494.
		\bibitem[Aneiros et al.(2019)Aneiros, Cao, Fraiman, Genest \& Vieu]{anecfgv19} ANEIROS, G., CAO, R., FRAIMAN, R., GENEST, C. \& VIEU, P. (2019). Recent advances in functional data analysis and high-dimensional statistics. \textit{Journal of Multivariate Analysis}   \pmb{170}, 3--9.
		\bibitem[Aneiros et al.(2015)Aneiros, Ferraty \& Vieu]{aneirosfv_2015} ANEIROS, G., FERRATY, F. \& VIEU, P. (2015). Variable selection in partial linear regression with functional covariate, \emph{Statistics}  \pmb{49}(6), 1322--1347.
		\bibitem[Aneiros \& Vieu(2015)]{aneirosv_2015} ANEIROS, G. \& VIEU, P. (2015). Partial linear modelling with multi-functional covariates.
		\emph{Computational Statistics} \pmb{30}(3), 647--671.
		\bibitem[Aneiros \& Vieu(2014)]{aneirosv_2014} ANEIROS, G. \& VIEU, P. (2014). Variable selection in infinite-dimensional problems.
		\emph{Statistics and Probability Letters} \pmb{94}, 12--20.
		\bibitem[Aneiros-P\'erez \& Vieu(2011)]{aneiros_2011} ANEIROS-P\'EREZ, G. \& VIEU, P. (2011). Automatic estimation procedure in  partial linear model with functional data. \emph{Statistical Papers} \pmb{52}(4), 751--771.
		\bibitem[Aneiros-P\'erez \& Vieu(2006)]{aneiros_2006} ANEIROS-P\'EREZ, G. \& VIEU, P. (2006). Semi-functional partial linear regression. \emph{Statistics and  Probability Letters} \pmb{76}, 1102--1110.
		
		
		\bibitem[Bakin(1999)]{bakin} BAKIN, S. (1999). Adaptive regression and model selection in data mining problems. PhD Thesis. Australian National University, Canberra.
		
		\bibitem[Carroll et al.(1997)Carroll, Fan, Gijbels  \& Wand]{carroll_1997}CARROL, R.J., FAN, J., GIJBELS, I. \& WAND, M. P. (1997), Generalized partially linear single-index models. \emph{Jounal of the American Statistical Association} \pmb{92}, 477–489.
		
		
		\bibitem[Fan \&  Li(2001)]{fan_li} FAN, J. \& LI, R. (2001). Variable selection via nonconcave penalized likelihood and its oracle properties. \emph{Journal of the American Statistical Association} \pmb{96}, 1348--1360.	
		\bibitem[Ferraty et al.(2003)Ferraty, Peuch \& Vieu]{ferpv03} FERRATY, F., PEUCH, A. \& VIEU, P. (2003). ``Mod\`{e}le \`{a} Indice Fonctionnel Simple'' (in French). \emph{Comptes Rendus Math\'ematique de l'Acad\'emie des Sciences Paris} \pmb{336}(12), 1025--1028.
		\bibitem[Ferraty \& Vieu(2006)]{fer2006} FERRATY, F. \&  VIEU, P. (2006).\emph{ Nonparametric Functional Data Analysis, Theory and Practice}. New York: Springer Series in Statistics.
		
		\bibitem[Goia \& Vieu(2016)]{goiv17} GOIA, A. \& VIEU, P. (2016). An introduction to recent advances in high/infinite dimensional statistics. \textit{Journal of Multivariate Analysis}  \pmb{146}, 1--6.
		
		
		\bibitem[Huand et al.(2008)Huang, Horowitz \& Ma]{huahowa} HUANG, J.  HOROWITZ, J.L. \& MA, S. (2008).  Asymptotic properties of bridge estimators in sparse high-dimensional regression models.
		\emph{The Annals of Statistics} \pmb{36}(2), 587--613.
		
		
		%\bibitem[Goia and Vieu(2014)]{goiv14} 
		%	Goia, A., Vieu, P. (2014). Some advances on semi-parametric functional data modelling. In Contributions in infinite-dimensional statistics and related topics.  Esculapio, Bologna, pp 135--140.
		\bibitem[Lian(2011)]{lian2011} LIAN, H. (2011). Functional partial linear model. \emph{Journal of Nonparametric Statistics} \pmb{23} (1), 115--128.
		\bibitem[Liang et al.(2010)]{liang_2010} LIANG, H., LIU, X.,  LI, R. \& TSAI, C. (2010). Estimation and Testing for Partially  Linear Single-Index Models. \emph{The Annals of Statistics} \pmb{38}(6), 3811–3836.
		\bibitem[Ling \& Vieu(2018)]{linv18}  LING, N. \& VIEU, P. (2018). Nonparametric modelling for functional data: selected survey and tracks for future. \emph{Statistics} \pmb{52}(4), 934--949.
		\bibitem[Maity \& Huang(2012)]{maity_2012} MAITY, A. \& HUANG, J. Z. (2012). Partially linear varying coefficient models stratified by a functional covariate. \emph{Statistics and Probability Letters} \pmb{82} (10), 1807--1814.
		\bibitem[Novo et al.(2019)Novo, Aneiros \& Vieu]{novo_2019} NOVO, S., ANEIROS, G. \& VIEU, P. (2019). Automatic and location-adaptive estimation in functional single-index regression. \emph{ Journal of Nonparametric Statistics} \pmb{31}(2), 364--392.
		\bibitem[Novo et al.(2020)Novo, Aneiros \& Vieu]{novo} NOVO, S., ANEIROS, G. \& VIEU, P. (2021). Sparse semiparametric regression when predictors are mixture of functional and high-dimensional variables. \emph{TEST} \pmb{30},  481--504.
		\bibitem[Shi et al.(2020)Shi, Huang, Jiao \& Yang]{shi} SHI, Y., HUANG, J., JIAO, Y. \& YANG, Q. (2020). A semismooth Newton algorithm for high-dimensional nonconvex sparse learning. \emph{IEEE Transactions on neural and learning systems} \pmb{31}(8), 2993--3006.
		%\bibitem[Shin(2009)]{shin2009}Shin, H. (2009), Partial Functional Linear Regression, \emph{Journal of Statistical Planning and Inference}, 139, 3405–3418.
		%\bibitem[Novo et al.(2019)]{novo_2019} Novo, S., Aneiros, G., Vieu, P. (2019). Sparse semiparametric regression when predictors are mixture of functional and high-dimensional variables.
		%	\bibitem[Aneiros-P\'erez and Vieu(2008)]{aneirosv_2008}  Aneiros-P\'erez, G., and Vieu P. (2008). Nonparametric time series prediction: a semi-functional partial linear modeling. \emph{Journal of Multivariate Analysis}, 99, 834-857.
		
		%	\bibitem[Fan and Peng(2001)]{fanpeng_2004}  Fan, J., and Peng, H. (2004). Nonconcave penalized likelihood with a diverging number of parameters. \emph{Journal of the American Statistical Association}, 96, 1348-1360.
		%	\bibitem[Ferraty et al.(2010)]{ferraty_2010}  Ferraty, F., Laksaci, A., Tadj, A., and Vieu, P. (2010). Rate of uniform consistency for nonparametric estimates with functional variables. \emph{Journal of Statistical Planning and Inference}, 140, 335-352.
		%	\bibitem[Ferraty and Vieu(2006)]{ferratyvieu_2006} Ferraty, F., and Vieu, P. (2006). \emph{Nonparametric Functional Data Analysis: Theory and Practice}. New York: Springer series in Statistics.
		
		\bibitem[Tibshirani(1996)]{tibshirani} TIBSHIRANI, R. (1996). Regression shrinkage and selection via the Lasso. {\it{Journal of the Royal Statistical Society, Series B}} \pmb{58}, 267--2888.
		
		\bibitem[Tibshirani \& Saunders(2005)]{tisa} TIBSHIRANI, R. \& SAUNDERS, M. (2005). Sparsity and smoothness via the fused lasso. {\it{Journal of the Royal Statistical Society, Series B}} \pmb{67}, 91--108.
		
		\bibitem[Vieu(2018)]{vie17} VIEU, P. (2018). On dimension reduction models for functional data. {\it{Statistics and  Probability  Letters}} \pmb{136}, 134--138.
		\bibitem[Wang et al.(2016)Wang, Feng \& Chen]{wang_2016} WANG, G., FENG, X.N. \& CHEN, M. (2016). Functional partial linear single-index model. \emph{Scandinavian Journal of Statistics}  \pmb{43}, 261-274.
		\bibitem[Zou(2006)]{zou} ZOU, H. (2006). The adaptive Lasso and its oracle properties. \emph{Journal of the American Statistical Association}  \pmb{101}, 1418--1429.
	\end{thebibliography}
\end{document}